\documentclass[12pt]{article}    
   \usepackage{latexsym} 
\usepackage{amssymb}  
\usepackage{amsfonts} 
\usepackage{multirow}
\usepackage{epsf}
\usepackage{epsfig}
\usepackage{rotate}

\def \lta {\mathrel{\vcenter
     {\hbox{$<$}\nointerlineskip\hbox{$\sim$}}}}

\renewcommand{\varphi}{\phi}

\newcommand{\Ga}{\Gamma}
\newcommand{\de}{\delta}
\newcommand{\De}{\Delta}

\newcommand{\la}{\lambda}

\newcommand{\dsp}{\displaystyle}

\newcommand{\beq}{\begin{equation}}
\newcommand{\eeq}{\end{equation}}
\newcommand{\ba}{\begin{array}}
\newcommand{\bea}{\begin{eqnarray}}
\newcommand{\ea}{\end{array}}
\newcommand{\eea}{\end{eqnarray}}

\newcommand{\ol}{\overline}

\newcommand\comment[1]{ \hbox{[{\it Comment suppressed here.}\/]} }
\newcommand\hide[1]{}

\newcommand{\skipover}[1]{}

\newcommand{\nnn} {\nonumber \vspace{.2cm} \\ }

\renewcommand{\O}{ {\cal O} }

\newcommand{\Tr}{\hbox{Tr}}

\def\ZZZ{Z\kern -0.31em Z}

\def\npb#1{Nucl. Phys. {\bf B#1}}

\def\prle#1{Phys. Rev. Lett. {\bf #1}}

\def\zpc#1{Z. Phys. {\bf C#1}}

\def\HPA#1{Helv. Phys. Acta~{\bf #1}}
\def\IJMP#1{Int. J. Mod. Phys.~{\bf #1}}

\def\MPL#1{Mod. Phys. Lett.~{\bf #1}}
\def\NP#1{Nucl. Phys.~{\bf #1}}

\def\PL#1{Phys. Lett.~{\bf #1}}
\def\PR#1{Phys. Rev.~{\bf #1}}
\def\PRP#1{Phys. Rep.~{\bf #1}}
\def\PRL#1{Phys. Rev. Lett.~{\bf #1}}
\def\PNAS#1{Proc. Nat. Acad. Sc.~{\bf #1}}

\def\ZP#1{Z. Phys.~{\bf #1}}

\def\appendix{\par                              
    \setcounter{section}{0}                     
    \setcounter{subsection}{0}
    \renewcommand{\theequation}{\Alph{section}.\arabic{equation}}
    \renewcommand{\thesection}{Appendix \Alph{section}
\setcounter{equation}{0}}
}

\catcode`\@=11

\def\applabel#1{\@bsphack
  \protected@write\@auxout{}%
         {\string\newlabel{#1}{{\Alph{section}}{\thepage}}}%
  \@esphack}

\def\section{
\setcounter{equation}{0}        
\@startsection {section}{1}{\z@}{-3.5ex plus -1ex minus 
 -.2ex}{2.3ex plus .2ex}{\large\bf}}
\renewcommand{\theequation}{\arabic{section}.\arabic{equation}}

\def\subsection{\@startsection{subsection}{2}{\z@}{-3.25ex plus -1ex minus 
 -.2ex}{1.5ex plus .2ex}{\normalsize\bf}}

\def\subsubsection{\@startsection{subsubsection}{3}{\z@}{-3.25ex plus
 -1ex minus -.2ex}{1.5ex plus .2ex}{\normalsize}}

\newcommand{\hal} {\frac{1}{2}}

\begin{document}
\setlength{\baselineskip}{16pt}

\title{\bf \Large Effective Average Action\\
in Statistical Physics and Quantum Field
Theory}

\author{Christof Wetterich\thanks{Email: 
{\tt C.Wetterich@thphys.uni-heidelberg.de}}\\
{\normalsize \it Institut f{\"u}r Theoretische Physik,
Universit{\"a}t Heidelberg}\\
{\normalsize \it 69120 Heidelberg, Germany}}

\date{}
\maketitle

\thispagestyle{empty}

\begin{abstract}
An exact renormalization group equation describes the dependence of the 
free energy on an infrared cutoff for the quantum or thermal fluctuations. 
It interpolates between the microphysical laws and the complex 
macroscopic phenomena. We present a simple unified description of 
critical phenomena for $O(N)$-symmetric scalar models in two, three or four dimensions, including 
essential scaling for the Kosterlitz-Thouless transition. \end{abstract}

\vspace*{\fill}

\setcounter{page}{1}


\newpage

\section{Introduction}

Particle physics and statistical physics share a common theoretical
challenge. For given ``microphysical laws'' observable ``macrophysical''
quantities at much larger characteristic length scales have to be
computed. This concerns, for example, the computation of the hadron
spectrum from perturbative QCD 
or the thermodynamic equation state for a liquid of helium from the 
atomic interactions.  The difficulties
arise from the importance of fluctuations. More technically, the
``transition to macrophysics''
requires the solution of a complicated functional integral. 
Beyond the identical conceptual setting of particle 
and statistical physics there is also quantitative agreement for certain
questions. The critical exponents at the endpoint of the critical
line of the electroweak phase transition (the onset of crossover
for the critical mass of the Higgs scalar)
are believed to be precisely the same as for the liquid-gas
transition near the critical pressure. This reflects universality.

Exact renormalization group equations  aim for a piecewise solution
of the fluctuation problem. They describe the scale dependence of
some type of ``effective action''. In this context an effective
action is a functional of fields from which the physical properties at a
given length or momentum scale can be computed. The exact
equations can be derived as formal identities from the
functional integral which defines the theory. They are cast in the
form of functional differential equations.
Different versions of exact renormalization
group equations have already a long history \cite{Kad66-1}--\cite{Has86-1}.
The investigation
of the generic features of their solutions has led to a deep
insight into the nature of renormalizability. In particle
physics, the discussion of perturbative solutions has led
to a new proof of perturbative renormalizability of the $\phi^4$-theory
\cite{Pol84-1}. Nevertheless, the application of exact renormalization group
methods to non-perturbative situations has been hindered for a long time 
by the complexity of functional differential equations. Some type of expansion is needed if one wants to
exploit the exactness of the functional differential equation
in practice -- otherwise any reasonable guess of a realistic
renormalization group equation does as well.

Since exact solutions to functional differential equations
seem only possible for some limiting
cases, it is crucial to find a formulation
which permits non-perturbative approximations. These proceed
by a truncation of the most general form of the effective
action and therefore need a qualitative understanding of
its properties. The formulation of an exact renormalization group
equation based on the effective average action \cite{Wet91-1,Wet93-2}
has been proven successful in this respect. It is the basis
of the non-perturbative flow equations which we discuss in this
review. The effective average action is the generating functional
of one-particle irreducible correlation functions in the presence of an
infrared cutoff scale $k$. Only fluctuations with momenta larger
than $k$ are included in its computation. For $k\to0$
all fluctuations are included and one
obtains the usual effective action from which appropriate masses and
vertices can be read off directly. The $k$ dependence is described
by an exact renormalization group equation that closely resembles
a renormalization group improved one-loop equation \cite{Wet93-2}. 
In fact, the
transition from classical propagators and vertices to effective
propagators and vertices transforms the one-loop expression into
an exact result. This close connection to perturbation theory for
which we have intuitive understanding is an important key for devising
non-perturbative approximations. Furthermore, the one-loop expression
is manifestly infrared and ultraviolet finite and can be used
directly in arbitrary number of dimensions, 
even in the presence of massless modes.

The aim of this review is to show that this version of flow equations
can be used in practice for the transition from microphysics
to macrophysics, even in the presence of strong couplings and a large
correlation length. It is largely
based on a more extensive report by
J. Berges, N. Tetradis, and the author \cite{report}. (We refer to this 
report as well as to \cite{BB} for a more extensive list of
references.)
We derive the exact renormalization group
equation and various non-perturbative truncations in section 
\ref{nonpertfloweq}. In
section 3 we discuss the solutions to the flow equation in more detail for $O(N)$-symmetric scalar models. We derive the 
scaling form of the flow equation for the 
effective potential and compute critical
exponents for second-order phase transitions in
three dimensions. Finally we discuss essential scaling
for the Kosterlitz-Thouless phase
transition for two-dimensional models with a continuous Abelian
symmetry.

Several results in statistical and particle physics have been
obtained first with the method presented here. This includes
the universal critical equation of state for spontaneous breaking of
a continuous symmetry in Heisenberg models \cite{BTW96-1}, 
the universal critical
equation of state for first-order phase transitions in matrix models 
\cite{BW97-1},
the non-universal critical amplitudes with an explicit connection
of the critical behavior to microphysics $(CO_2)$ \cite{B}, a
quantitatively reliable estimate of the rate of spontaneous
nucleation \cite{first,second}, 
a classification of all possible fixed points for
(one component) scalar theories in two and three dimensions in
case of a weak momentum dependence of the interactions 
\cite{Mortwod},
the second-order phase transition in the high temperature
$\phi^4$ quantum field theory \cite{TW93-1}, 
the phase diagram for the Abelian
Higgs model for $N$ charged scalar fields \cite{BLL95-1,Tet96-1}.
In particle physics it has led to  
the prediction that the
electroweak interactions become strong at high temperature,
with the suggestion that the standard model may show a crossover
instead of a phase transition \cite{RW93-2,W-Sintra}. 
In strong interaction physics
the interpolation between the low temperature chiral perturbation
theory results and the high temperature critical behavior for 
two light quarks has been established
in an effective model \cite{BJW97-1}.  
All these results are in the non-perturbative domain. In addition,
the approach has been tested successfully through a comparison with
known high precision results on critical exponents and universal
critical amplitude ratios and effective couplings.

Our main conclusion can be drawn already at this place: the
method works in practice. If needed and with sufficient
effort  precise results can be obtained in many
non-perturbative situations. New phenomena become accessible 
to a reliable analytical treatment.

\section{Non-Perturbative flow equation}
\label{nonpertfloweq}

\subsection{Average action}
\label{AverAct}

We will concentrate on a flow equation
which describes the scale dependence of the effective
{average action} $\Gamma_k$ \cite{Wet93-2}. The latter
is based on the quantum field theoretical concept of the effective
action $\Gamma$, i.e.\ the generating functional of
the Euclidean one-particle irreducible ($1 PI$) correlation functions
or proper vertices. 
For the vacuum or ground state this
functional is obtained after ``integrating
out'' the quantum fluctuations. The masses of the excitations,
scattering amplitudes and cross
sections follow directly from an analytic continuation of the 1PI
correlation functions in a standard way. Furthermore, the
field equations derived from the
{\em effective action} are exact as all quantum effects are
included.
In thermal and chemical equilibrium
$\Gamma$ includes in addition the thermal
fluctuations and depends on the temperature $T$ and chemical potential
$\mu$.
In statistical physics $\Gamma$ is related to
the free energy as a functional
of some space-dependent order parameter $\phi(x)$. For vanishing
external fields the equilibrium state is given by the
minimum of $\Gamma$. More generally, in the presence of (spatially
varying) external fields or sources the equilibrium state obeys
\beq\label{2.1}
\frac{\delta\Gamma}{\delta\phi(x)}=J(x),\eeq
and the precise relation to the thermodynamic potentials like the
free energy $F$ reads
\beq
\label{2.2}
F=T\Gamma_{eq}+\mu \overline N-T\int dx \, \phi_{eq}(x)J(x).
\eeq
Here $\phi_{eq}(x)$ solves (\ref{2.1}), $\Gamma_{eq}=\Gamma[\phi
_{eq}]$, and $\overline N$ is the conserved quantity to which the chemical potential
is associated. For homogeneous $J=j/T$ 
the equilibrium
value of the order parameter $\phi$ is often also homogeneous.
In this case the energy density $\epsilon$, entropy density $s$,
``particle density'' $n$ and pressure $p$ can be simply expressed
in terms of the effective potential $U(\phi)=T\Gamma/V$, namely
\bea\label{2.3}
&&\epsilon=U-T\frac{\partial U}{\partial T}-\mu\frac{\partial U}
{\partial\mu}\ ,\ s=-\frac{\partial U}{\partial T}+\frac{j\phi}{T},
\nonumber\\
&&n=-\frac{\partial U}{\partial\mu}\ ,\ p=-U=-T\Gamma/V.
\eea
Here $U$ has to be evaluated for the solution of $\partial U/\partial
\phi=j$, $n=\overline N/V$ and $V$ is the total volume of (three-dimensional)
space. Evaluating
$U$ for arbitrary $\phi$ yields the equation of state in the presence
of homogeneous magnetic fields or other appropriate 
sources.

More formally, the effective action $\Gamma$ follows from a Legendre
transform of the logarithm of the partition function in the presence of
external sources or fields (see below). Knowledge of $\Gamma$ is
in a sense equivalent to the ``solution'' of a theory. Therefore
$\Gamma$ is the
macroscopic quantity on which we will concentrate. In particular,
the effective potential $U$ contains already a large part of
the macroscopic information relevant for homogeneous states. We
emphasize that the concept of the effective potential
is valid universally for classical
statistics and quantum statistics, or quantum field theory in
thermal equilibrium or the vacuum\footnote{The only difference
concerns the evaluation of the partition function $Z$ or $W=\ln Z=
-(F-\mu N)/T$. For classical statistics it involves a $D$-dimensional
functional integral, whereas for quantum statistics the dimension
in the Matsubara formalism is $D+1$. The vacuum in quantum field
theory corresponds to $T\to0$, with $V/T$ the volume of Euclidean
``spacetime''.}.

The average action $\Gamma_k$ is a simple
generalization of the effective action, with the distinction that only
fluctuations with momenta $q^2 \gtrsim k^2$ are included.
This is achieved
by implementing an infrared (IR) cutoff $\sim k$ in the functional
integral that defines the effective action $\Gamma$. In the
language of statistical physics, $\Gamma_k$ is a type of
coarse-grained free
energy with a coarse graining length scale $\sim k^{-1}$. As long as
$k$ remains large enough, the possible complicated effects of
coherent long-distance fluctuations play no role and $\Gamma_k$ is close
to the microscopic action.
Lowering $k$ results in
a successive inclusion of fluctuations with momenta
$q^2 \gtrsim k^2$ and therefore permits to explore the theory on
larger and larger length scales. The average action $\Gamma_k$ can be viewed
as the effective action for averages of fields over a volume
with size $k^{-d}$ \cite{Wet91-1} 
and is similar in spirit to the action for block--spins
on the sites of a coarse lattice.

By definition, the average action equals the standard effective action
for $k=0$, i.e.\ $\Gamma_{0}=\Gamma$, as the IR cutoff is
absent in this limit
and all fluctuations are included. On the other hand,
in a model with a physical
ultraviolet (UV) cutoff $\Lambda$ we
can associate $\Gamma_\Lambda$ with the microscopic or
classical action $S$.
No fluctuations with momenta below $\Lambda$ are effectively included
if the IR cutoff equals the UV cutoff.
Thus the average action $\Gamma_k$ has the important property that
it interpolates between the classical
action $S$
and the effective action $\Gamma$ as $k$ is lowered from the ultraviolet
cutoff $\Lambda$ to zero:
\beq\label{2.4}
\Gamma_\Lambda \approx S\ ,
\,\, \lim\limits_{k \to 0} \Gamma_k = \Gamma.\eeq
The ability to follow the evolution to $k\rightarrow0$ is equivalent to the
ability to solve the theory. Most importantly, the dependence of the
average action on the scale $k$ is described by an exact
non-perturbative flow equation presented in the next 
subsection.\\

Let us consider the construction of $\Gamma_k$ for a
simple model with real scalar fields $\chi_a$, $a=1 \ldots N$, in $d$
Euclidean dimensions with classical action $S$.
We start with the path integral representation of the
generating functional for the connected correlation functions in
the presence of an IR cutoff. It is given by the logarithm of the
(grand) canonical partition function in the presence of inhomogeneous
external fields or sources $J_{a}$
\beq\label{2.5}
W_k[J]=\ln Z[J]=\ln \int D \chi \exp\left(-S_{}[\chi]-\De S_k[\chi]+
\int d^dx J_a(x)\chi^a(x) \right)\; .
\label{genfunc}
\eeq
In classical statistical physics $S$ is related to the Hamiltonean
$H$ by $S=H/T$, so that $e^{-S}$ is the usual Boltzmann factor.
The functional integration $\int D\chi$ stands for the sum
over all microscopic states. In turn, the field $\chi_a(x)$ can
represent a large variety of physical objects like a (mass-)
density field $(N=1)$, a local magnetisation $(N=3)$ or
a charged order parameter $(N=2)$. The only modification
compared to the construction of the standard effective action
is the addition of an
IR cutoff term $\De S_k[\chi]$ which is
quadratic in the fields and reads
in momentum space $(\chi_a(-q)\equiv\chi_a^*(q))$
\beq\label{2.6}
\De S_k[\chi]=\hal \int
\frac{d^dq}{(2\pi)^d} R_k(q)\chi_a(-q)\chi^a(q).
\eeq
Here the IR cutoff function $R_k$ is required to vanish
for $k \to 0$ and to diverge for $k \to \infty$ (or $k\to\Lambda)$
and fixed $q^2$.
This can be achieved, for example, by the
exponential form
\beq\label{2.7}
 R_k(q) = \frac{Z_kq^2}{\displaystyle{e^{q^2/k^2} - 1}}
 \label{Rk(q)}
\eeq
where the wave function renormalization constant $Z_k$ will
be defined later.
For fluctuations with small momenta
$q^2\ll k^2$ this cutoff behaves as $R_k(q)\sim k^2$ and allows
for a simple interpretation: Since
$\De S_k[\chi]$ is quadratic in the fields,
all Fourier modes of $\chi$ with momenta smaller than
$k$ acquire an effective mass $\sim k$. This additional mass term
acts as an effective IR cutoff for the low momentum modes.
In contrast, for $q^2\gg k^2$ the function $R_k(q)$ vanishes
so that the functional integration of the high momentum modes
is not disturbed. The term $\De S_k[\chi]$ added to the classical
action is the main ingredient for the construction of an effective
action that includes all fluctuations with momenta $q^2 \gtrsim k^2$,
whereas fluctuations with $q^2 \lesssim k^2$ are suppressed.

The expectation value of $\chi$, i.e.\ the macroscopic field $\phi$,
in the presence of $\De S_k[\chi]$ and $J$ reads
\beq
\label{2.8}
\phi^a(x) \equiv \langle\chi^a(x)\rangle =
\frac{\de W_k[J]}{\de J_a(x)}.
\eeq
We note that the relation between $\phi$ and $J$ is
$k$--dependent, $\phi=\phi_k[J]$ and therefore \mbox{$J=J_k[\phi]$}.
In terms of $W_k$ the average action is
defined via a modified Legendre transform
\beq
\label{2.9}
\Ga_k[\phi]=-W_k[J]+\int d^dx J_a(x)\phi^a(x)-\De S_k[\phi],
\eeq
where we have subtracted the term $\De S_k[\phi]$ in the rhs.
This subtraction of the IR cutoff term as a function of the
macroscopic field $\phi$ is crucial for the definition of a
reasonable coarse-grained free energy with the property
$\Gamma_\Lambda\approx S$. It guarantees
that the only difference between $\Gamma_k$ and $\Gamma$ is the
effective infrared cutoff in the fluctuations.
Furthermore, it has the consequence that $\Gamma_k$ does not
need to be convex, whereas a pure Legendre transform is always convex by
definition. The coarse-grained free energy has to
become convex \cite{RingWet90,TetWet92}
only for $k\rightarrow0$. These considerations are important for an
understanding of spontaneous symmetry breaking and, in particular,
for a discussion of nucleation in a first-order phase transition.

In order to establish the property $\Gamma_{\Lambda}\approx S$ we consider an
integral equation for $\Gamma_k$ that is equivalent to (\ref{2.9}).
In an obvious matrix notation, where
$J\chi\equiv\int d^dxJ_a(x)\chi^a(x)=
\int\frac{d^dp}{(2\pi)^d}J_a(-p)\chi_a(p)$
and $R_{k,ab}(q,q')=R_k(q)\delta_{ab}(2\pi)^d\delta(q-q')$, we represent
(\ref{2.5}) as
\beq\label{2.10}
\exp\Big(W_k[J]\Big)= \int D \chi \exp\left(-S[\chi]+ J \chi
-\frac{1}{2} \chi\, R_k \chi \right)\; .
\eeq
As usual, we can invert the Legendre transform (\ref{2.9}) to express
\beq\label{2.11}
J=\frac{\de \Gamma_k}{\de \phi}+\phi\, R_k.
\eeq
It is now straightforward to insert the definition
(\ref{2.9}) into (\ref{2.10}). After a variable substitution
$\chi'=\chi-\phi$ one obtains the functional integral representation
of $\Gamma_k$
\beq
\label{2.12}
\exp(-\Gamma_k[\phi])=\dsp{\int D \chi' \exp\left(-S[\phi+\chi']
+ \frac{\de \Gamma_k}{\de \phi}
\chi'  -\frac{1}{2}
\chi' \, R_k\, \chi'  \right)}
\; .\eeq
This expression resembles closely the background field formalism
for the effective action which is modified only by the term $\sim
R_k$.
For $k \to \infty$ the cutoff function $R_k$ diverges and
the term $\exp( -\chi'  R_k \chi' /2) $ behaves
as a delta functional $\sim \de[\chi']$, thus leading to the
property $\Gamma_k \to S$
in this limit. For a model with a sharp UV cutoff $\Lambda$ it is
easy to enforce the identity $\Gamma_\Lambda=S$ by choosing a cutoff
function $R_k$ which diverges for $k\to\Lambda$, like
$R_k\sim q^2(e^{q^2/k^2}-e^{q^2/\Lambda^2})^{-1}$.
We note, however,
that the property $\Gamma_\Lambda=S$ is not essential, as the short
distance laws may be parameterized by $\Gamma_\Lambda$ as
well as by $S$. For momentum scales much smaller than
$\Lambda$ universality implies
that the precise form of $\Gamma_\Lambda$ is irrelevant,
up to the values of a
few relevant renormalized couplings.
Furthermore, the microscopic action may be formulated on a lattice
instead of continuous space and can involve even variables different
from $\chi_a(x)$. In this case one can still compute $\Gamma_\Lambda$
in a first step by evaluating the functional integral (\ref{2.12})
approximately. Often a saddle point expansion will do, since no
long-range fluctuations are involved in the
transition  from $S$ to $\Gamma_\Lambda$. In
this report we will assume that the first step of the computation
of $\Gamma_\Lambda$ is done and consider $\Gamma_\Lambda$ as the
appropriate parametrization of the microscopic physical laws.
Our aim is the computation of the effective action $\Gamma$
from $\Gamma_\Lambda$ -- this step may be called ``transition
to complexity'' and involves fluctuations on all scales. We emphasize
that for large $\Lambda$ the average action $\Gamma_\Lambda$
can serve as a formulation of the microscopic laws also for situations
where no physical cutoff is present, or where a momentum UV cutoff
may even be in conflict with the symmetries, like the important
case of gauge symmetries.

A few properties of the effective average action are worth mentioning:
\begin{enumerate}
\item All symmetries of the model which are respected by the IR cutoff
  $\Delta S_k$ are automatically symmetries of $\Gamma_k$. In
  particular this concerns translation and rotation invariance. As a result
  the approach is not plagued by many of the problems encountered by a
  formulation of the block-spin action on a lattice. 
  Nevertheless, our method is not restricted to continuous space.
  For a cubic lattice with lattice distance $a$ the propagator 
  only obeys the restricted lattice translation and rotation
  symmetries, e.g.\ a next neighbor interaction leads in momentum
  space to 
\beq
S=\frac{2}{a^2} \int \frac{d^dq}{(2 \pi)^d} \sum\limits_{\mu}
\left(1-\cos a q_{\mu}\right) \chi^*(q) \chi(q) + \ldots 
\eeq  
  The momentum cutoff $|q_{\mu}|\le \Lambda$, $\Lambda = \pi / a$
  also reflects the lattice symmetry. A rotation and translation
  symmetric cutoff $R_k$ which only depends on $q^2$ obeys 
  automatically all possible lattice symmetries. The only 
  change compared to continuous space will be the reduced symmetry
  of $\Gamma_k$.
\item In consequence, $\Gamma_k$ can be expanded in terms of
  invariants with respect to the symmetries with couplings depending
  on $k$. For the example of a scalar $O(N)$-model in continuous space
  one may use a derivative
  expansion ($\rho=\phi^a\phi_a/2$)
  \begin{equation}
    \label{2.13}
    \Gamma_k=\int d^d x\left\{
    U_k(\rho)+\frac{1}{2}Z_{\phi,k}(\rho)
    \partial^\mu\phi_a\partial_\mu\phi^a+\ldots\right\}
  \end{equation}
  and expand further in powers of $\rho$
  \begin{eqnarray}
    \label{2.14}
    \dsp{U_k(\rho)} &=& \dsp{
    \frac{1}{2}\ol{\la}_k\left(
    \rho-\rho_0(k)\right)^2+
    \frac{1}{6}\ol{\gamma}_k\left(
    \rho-\rho_0(k)\right)^3+\ldots}\nnn
    \dsp{Z_{\phi,k}(\rho)} &=& \dsp{
    Z_{\phi,k}(\rho_0)+Z_{\phi,k}^\prime(\rho_0)
    \left(\rho-\rho_0\right)+\ldots}\; 
  \end{eqnarray}
  Here $\rho_0$ denotes the ($k$--dependent) minimum of the effective average
  potential $U_k(\rho)$.  We see that $\Gamma_k$ describes infinitely many
  running couplings.
\item Up to an overall scale factor, the limit $k\to 0$ of $U_k$
corresponds to the effective potential $U=T\Gamma /V$, from
which the thermodynamic quantities can be derived for homogeneous
situations according to eq.\ (\ref{2.3}).
The overall scale factor is fixed by dimensional considerations.
Whereas the dimension of $U_k$ is (mass)$^d$ the dimension of
$U$ in eq.\ (\ref{2.3}) is (mass)$^4$ (for $h\!\!\bar{}\,\,=c=k_B=1$).
For classical statistics in $d=3$ dimensions one has
$U_k=\Gamma_k/V$ and $U=T\lim_{k\to 0}U_k$. For two-dimensional
systems an additional factor $\sim$ mass appears, since
$U_k=\Gamma_k/V_2=L \Gamma_k/V$ implies
$U=T L^{-1} \lim_{k\to 0} U_k$. Here  $L$ is the typical thickness
of the two-dimensional layers in a physical system.
In the following we will often omit these scale factors.
\item The functional $\tilde{\Gamma}_k[\phi]=\Gamma_k[\phi]
+\Delta S_k[\phi]$ is the Legendre transform of $W_k$ and 
therefore convex. This implies that all eigenvalues 
of the matrix of second functional derivatives $\Gamma^{(2)}+R_k$
are positive semi-definite. In particular, one finds for
a homogeneous field $\phi_a$ and $q^2=0$ the simple exact bounds for
all $k$ and $\rho$
\bea
\label{PosDef}
U_k^{\prime}(\rho) &\ge& -R_k(0) \nnn
U_k^{\prime}(\rho) + 2 \rho U_k^{\prime\prime}(\rho) &\ge& -R_k(0),
\eea
where primes denote derivatives with respect to $\rho$.
Even though the potential $U(\phi)$ becomes convex for $k\to 0$
it may exhibit a minimum at $\rho_0(k) >0$ for all $k>0$.
Spontaneous breaking of the $O(N)$-symmetry is characterized by
$\lim_{k\to 0} \rho_0(k) > 0$.
\item The results for physical quantities have to be independent
  of the choice of the cutoff scheme $R_k$. On the other hand,
  both $\Gamma_\Lambda$ and the flow with $k$ are scheme-dependent.
  The scheme independence of the final results is a good check for
  approximations
  \cite{Wet91-1,BHLM95-1,LitimS,LPS00,SSATM}. 
\item There is no problem incorporating chiral fermions, since a chirally
  invariant cutoff $R_k$ can be formulated \cite{Wet90-1,CKM97-1}. Gauge
  theories can be formulated along similar
  lines\footnote{See also \cite{gravity} for applications to gravity. } 
  \cite{28a,RW93-1,RW93-2}, \cite{Bec96-1}--\cite{FLP}
  even though $\Delta S_k$ may not be gauge
  invariant\footnote{For a manifestly gauge invariant formulation in terms
  of Wilson loops see ref. \cite{Mor98-1}.}. In this case the usual Ward
  identities receive corrections\footnote{The
  background field identity derived in \cite{28a}
  is equivalent \cite{FrW,FLP} to the modified
  Ward identity.} for which one can derive closed
  expressions \cite{28a,EHW94-1}. These corrections vanish for
  $k\rightarrow0$. On the other hand they appear as ``counterterms''
  in $\Gamma_\Lambda$ and are crucial for preserving the gauge
  invariance of physical quantities. 
\item Despite a similar spirit and many analogies, there is a
  conceptual difference to the Wilsonian effective action
  $S_\Lambda^{\rm W}$.
  The Wilsonian effective action describes a set of different actions
  (parameterized by $\Lambda$) for one and the same
  model --- the n--point
  functions are independent of $\Lambda$ and have to be computed from
  $S_\Lambda^{\rm W}$ by further functional integration. In contrast,
  $\Gamma_k$ can be viewed as the effective action for
  a set of  different
  ``models'' --- for any
  value of $k$ the effective average action is related to the generating
  functional of 1PI $n$-point functions for
  a model with a different action $S_k=S+\Delta S_k$. The
  $n$--point functions depend on $k$. The Wilsonian effective action does
  not generate the 1PI Green functions \cite{KKS92-1}.
\end{enumerate}

\subsection{Exact flow equation}
\label{ExactFlow}

The dependence of the average action $\Gamma_k$ on the
coarse graining scale $k$ is
described by an exact non-perturbative flow equation 
\cite{Wet93-2,BAM,Mor94,El}
\begin{equation}\label{2.17}
  \frac{\partial}{\partial k}\Ga_k[\phi] =  \hal\Tr\left\{\left[
  \Ga_k^{(2)}[\phi]+R_k\right]^{-1}\frac{\partial}{\partial k}
  R_k\right\} \; .
\end{equation}
The trace involves an integration over momenta or coordinates
as well as a summation over internal
indices. In momentum space it reads
$\Tr=\sum_a \int d^dq/(2\pi)^d$, as appropriate for the unit matrix
${\bf 1}=(2 \pi)^d \delta(q-q^\prime )\delta_{ab}$. 
The exact flow equation describes
the scale dependence of $\Gamma_k$ in terms of the inverse
average propagator $\Ga_k^{(2)}$, given by
the second functional derivative of $\Ga_k$ with respect
to the field components
\begin{equation}\label{2.18}
  \left(\Gamma_k^{(2)}\right)_{a b}(q,q^\prime)=
  \frac{\delta^2\Gamma_k}
  {\delta\phi^a(-q)\delta\phi^b(q^\prime)}\; .
\end{equation}
It has a simple graphical expression as a one-loop equation 

\vspace*{0.2in}
\hspace*{5.5cm}
$\displaystyle{\frac{\partial \Gamma_k}
{\partial k}\,\,\,\, =\,\,\,\,\frac{1}{2}\,\,\,\,}$
\parbox{1.in}{
\epsfxsize=.4in
\epsffile{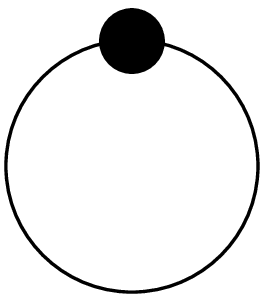}}
\vspace*{0.2in}

\noindent
with the full $k$--dependent propagator associated to the propagator
line and the dot denoting the insertion $\partial_k R_k$.

Because of the appearance
of the exact propagator $(\Gamma_k^{(2)}+R_k)^{-1}$, eq. (\ref{2.17}) is
a functional differential equation. It is remarkable that the
transition from the classical propagator in the presence of the
infrared cutoff, $(S^{(2)}+R_k)^{-1}$, to
the full propagator turns the one-loop expression into an exact
identity which incorporates effects of arbitrarily high loop order
as well as genuinely non-perturbative effects\footnote{We note
that anomalies which arise from topological obstructions in
the functional measure manifest themselves already in the
microscopic action $\Gamma_\Lambda$. The long-distance
non-perturbative effects (``large-size instantons'') are,
however, completely described by the flow equation (\ref{2.17}).}
like instantons in QCD.

The exact flow equation (\ref{2.17}) can be derived in a
straightforward way \cite{Wet93-2}. Let us write
\beq\label{2.19}
\Gamma_k[\phi]=\tilde{\Gamma}_k[\phi]-\Delta S_k[\phi],
\eeq
where, according to (\ref{2.9}),
\beq\label{2.20}
\tilde{\Gamma}_k[\phi]=-W_k[J]+\int d^dx J(x)\phi(x)
\eeq
and $J=J_k(\phi)$. We consider for simplicity a one--component
field and derive first the scale dependence of $\tilde{\Gamma}$:
\beq\label{2.21}
\frac{\partial}{\partial k} \tilde{\Gamma}_k[\phi] =
-\left(\frac{\partial W_k}{\partial k}  \right) [J]
-\int d^dx \frac{\delta W_k}{\delta J(x)}
\frac{\partial J(x)}{\partial k} + \int d^dx \phi(x)
\frac{\partial J(x)}{\partial k} \; .
\eeq
With $\phi(x)=\delta W_k/\delta J(x)$ the last two terms
in (\ref{2.21}) cancel. The $k$--derivative
of $W_k$ is obtained from its defining functional integral
(\ref{2.5}). Since only $R_k$ depends on $k$ this
yields
\beq\label{2.22}
\frac{\partial}{\partial k} \tilde{\Gamma}_k[\phi]=
\left\langle \frac{\partial}{\partial k}\Delta S_k[\chi] \right\rangle =
\left\langle \frac{1}{2} \int d^dx d^dy \chi(x)
\frac{\partial}{\partial k}R_k(x,y) \chi(y) \right\rangle \; ,
\eeq
where $R_k(x,y)=R_k(x-y)$ is the Fourier
transform of $R_k(q)$ and
\beq\label{2.23}
\left\langle A[\chi] \right\rangle =Z^{-1}\int D\chi A[\chi]\exp(-S[\chi]-
\Delta_kS[\chi]+\int d^dxJ(x)\chi(x)).\eeq
Let
$G(x,y)=\delta^2 W_k/\delta J(x)\delta J(y)$
denote the connected $2$--point function and decompose
\beq\label{2.24}
\langle \chi(x) \chi(y) \rangle = G(x,y) + \langle \chi(x) \rangle
\langle \chi(y) \rangle \equiv G(x,y) + \phi(x) \phi(y) \; .
\eeq
After plugging this decomposition into (\ref{2.22}), the scale
dependence of $\tilde{\Gamma}_k$ can be expressed as
\bea\label{2.25}
\frac{\partial}{\partial k} \tilde{\Gamma}_k[\phi] &=&
\frac{1}{2} \int d^dx d^dy \left\{   \frac{\partial}{\partial k}
R_k(x,y)G(y,x)
+ \phi(x) \frac{\partial}{\partial k} R_k(x,y) \phi(y) \right\} \nnn
&\equiv&
\frac{1}{2} \Tr\left\{ G \frac{\partial}{\partial k} R_k\right\}
+\frac{\partial}{\partial k} \Delta S_k[\phi] \; .
\eea
The exact flow equation for the average action $\Gamma_k$
follows now through (\ref{2.19})
\beq\label{2.26}
\frac{\partial}{\partial k} \Gamma_k[\phi] =
\frac{1}{2} \Tr\left\{ G \frac{\partial}{\partial k} R_k\right\}
= \frac{1}{2} \Tr\left\{ \left[
  \Ga_k^{(2)}[\phi]+R_k\right]^{-1} \frac{\partial}{\partial k} R_k
\right\}.
\eeq
For the last equation we have used that
$\tilde{\Gamma}_k^{(2)}(x,y)\equiv
\delta^2\tilde{\Gamma}_k/\delta\phi(x)\delta\phi(y)=
\delta J(x)/\delta \phi(y)$
is the inverse of $G(x,y)\equiv \delta^2 W_k/\delta J(x)\delta J(y)=
\delta \phi(x)/\delta J(y)$:
\beq\label{2.26a}
\int d^dyG(x,y)\left( \Gamma_k^{(2)}+R_k \right)(y,z)=\delta(x-z)\eeq
It is straightforward to write
the above identities in momentum space and to generalize them to
$N$ components by using the matrix notation introduced above.

Let us point out a few properties of the exact flow equation:
\begin{enumerate}
\item For a scaling form of the evolution equation and a formulation
closer to the usual $\beta$-functions, one may replace the partial
$k$-derivative in (\ref{2.17}) by a partial derivative with
respect to the logarithmic variable $t=\ln (k/\Lambda)$.

\item Exact flow equations for $n$--point functions can be easily obtained
from (\ref{2.17}) by differentiation. The flow equation for the
two--point function $\Gamma_k^{(2)}$ involves the three and
four--point functions, $\Ga_k^{(3)}$ and $\Ga_k^{(4)}$, respectively.
One may write schematically
\bea\label{2.28}
\frac{\partial}{\partial t} \Gamma_k^{(2)} &=&
\frac{\partial}{\partial t} \frac{\partial^2\Gamma_k}
  {\partial\phi\partial\phi}\nnn
&=& -\frac{1}{2} \Tr \left\{ \frac{\partial R_k}{\partial t}
\frac{\partial}{\partial \phi}
\left(\left[\Ga_k^{(2)}+R_k\right]^{-1} \Ga_k^{(3)}
\left[\Ga_k^{(2)}+R_k\right]^{-1}\right)\right\}\nnn
&=& \Tr \left\{ \frac{\partial R_k}{\partial t}
\left[\Ga_k^{(2)}+R_k\right]^{-1} \Ga_k^{(3)}
\left[\Ga_k^{(2)}+R_k\right]^{-1} \Ga_k^{(3)}
\left[\Ga_k^{(2)}+R_k\right]^{-1}\right\}\nnn
&-&\frac{1}{2} \Tr \left\{ \frac{\partial R_k}{\partial t}
\left[\Ga_k^{(2)}+R_k\right]^{-1} \Ga_k^{(4)}
\left[\Ga_k^{(2)}+R_k\right]^{-1}\right\} \; .
\eea
By evaluating this equation for $\phi=0$, one sees immediately
the contributions to the flow of the two-point function
from diagrams with three- and four-point vertices. 
The diagramatics
is closely linked to the perturbative graphs.
In general, the flow equation for $\Ga_k^{(n)}$ involves
$\Ga_k^{(n+1)}$ and $\Ga_k^{(n+2)}$.

\item As already mentioned, the flow
equation (\ref{2.17}) closely resembles a one--loop equation.
Replacing $\Gamma_k^{(2)}$ by the second functional derivative of the
classical action, $S^{(2)}$, one obtains the corresponding one--loop
result. Indeed, the one--loop formula for
$\Gamma_k$ reads
\begin{equation}
  \label{2.27}
  \Gamma_k[\phi]=S[\phi]+
  \frac{1}{2}\Tr\ln\left(
  S^{(2)}[\phi]+R_k\right)
\end{equation}
and taking a $k$--derivative of
(\ref{2.27}) gives a one--loop flow equation very similar to
(\ref{2.17}). The ``full renormalization
group improvement'' $S^{(2)}\rightarrow\Gamma_k^{(2)}$ turns the one--loop
flow equation into an exact non-perturbative flow equation.
Replacing the propagator and vertices appearing in
$\Gamma_k^{(2)}$ by the ones derived from the classical
action, but with running $k$--dependent couplings, and expanding
the result to lowest non--trivial order in the coupling constants, one
recovers standard renormalization group improved one--loop
perturbation theory.

\item The additional cutoff function $R_k$ with a form like
the one given in eq.\ (\ref{2.14}) renders the momentum integration
implied in the trace of (\ref{2.17}) both
infrared and ultraviolet finite. In particular, for $q^2\ll k^2$
one has an additional
mass--like term $R_k \sim k^2$
in the inverse average propagator. This makes the formulation suitable
for dealing with theories that are plagued with infrared problems
in perturbation theory.
For example, the flow equation can be used in three dimensions
in the phase with spontaneous symmetry breaking despite the
existence of massless Goldstone bosons for $N>1$. We recall
that all eigenvalues of the matrix $\Gamma^{(2)}+R_k$ must be
positive semi-definite (cf.\ eq.\ (\ref{PosDef})). We note that
the derivation of the exact flow equation does not
depend on the particular choice of the cutoff function.
Ultraviolet finiteness, however, is related
to a fast decay of $\partial_t R_k$ for $q^2\gg k^2$.
If for some other
choice of $R_k$ the rhs of the flow equation would not
remain ultraviolet finite this would indicate
that the high momentum modes have
not yet been integrated out completely in the computation of
$\Gamma_k$. Unless stated otherwise we will always assume a
sufficiently fast decaying choice of $R_k$ in the following.

\item Since no infinities appear in the flow equation, one may
``forget'' its origin in a functional integral. Indeed,
for a given choice of the cutoff function $R_k$ all microscopic
physics is encoded in the microscopic effective action $\Gamma_\Lambda$.
The model is completely specified by the flow equation (\ref{2.17})
and the ``initial value'' $\Gamma_\Lambda$. In a quantum field
theoretical sense the flow equation defines a regularization
scheme. The ``ERGE''-scheme is specified by the flow equation,
the choice of $R_k$ and the ``initial condition'' $\Gamma_\Lambda$.
This is particularly important for gauge theories where other
regularizations in four dimensions and in the presence of chiral
fermions are difficult to construct. For gauge theories $\Gamma_\Lambda$
has to obey appropriately modified Ward identities. In the context
of perturbation theory a first proposal for how to regularize gauge
theories by use of flow equations can be found in \cite{Bec96-1}. We note that
in contrast to previous versions of exact renormalization group
equations there is no need in the present formulation to construct
an ultraviolet momentum cutoff -- a task known to be
very difficult in non-Abelian gauge theories.

\item
Extensions of the
flow equations to gauge fields \cite{RW93-1,RW93-2}, \cite{Bec96-1}--\cite{FLP}
and fermions~\cite{Wet90-1,CKM97-1} are available.

\item
We emphasize that the flow equation (\ref{2.17}) is 
mathematically equivalent
to the Wilsonian exact renormalization group
equation \cite{Wil71-1,WH73-1,NC77,Wei76-1,Pol84-1,Has86-1}.
The latter describes how
the Wilsonian effective action $S_\Lambda^{\rm W}$ changes with an
ultraviolet cutoff $\Lambda$. Polchinski's continuum
version of the Wilsonian flow
equation~\cite{Pol84-1} can be transformed into eq.\ (\ref{2.17})
by means of a Legendre transform,  a suitable field
redefinition and the association $\Lambda=k$~\cite{BAM,ELL93,WetIJMP94}.
Although the formal relation is simple, the practical calculation
of $S^W_k$ from $\Gamma_k$ (and vice versa)
can be quite involved\footnote{If
this problem could be solved, one would be able to construct
an UV momentum cutoff which preserves gauge invariance by
starting from the Ward identities for $\Gamma_k$.}. In the presence
of massless particles the Legendre transform of $\Gamma_k$
does not remain local and $S^W_k$ is a comparatively
complicated object. We will argue below that the crucial
step for a practical use of the flow equation in a non-perturbative
context is the ability to device a reasonable approximation
scheme or truncation. It is in this context that the close
resemblence of eq. (\ref{2.17}) to a perturbative expression is
of great value.

\item
In contrast to the Wilsonian effective action no information
about the short-distance physics is effectively lost as $k$ is
lowered. Indeed, the effective average action for fields with high
momenta $q^2\gg k^2$ is already very close to the effective action.
Therefore $\Gamma_k$ generates quite accurately the vertices
with high external momenta. More precisely, this is the case
whenever the external momenta act effectively as an independent
``physical'' IR cutoff in the flow equation for the vertex. There
is then only a minor difference between $\Gamma_k^{(n)}$ and the
exact vertex $\Gamma^{(n)}$.

\item
An exact equation of the type (\ref{2.17}) can be derived
whenever $R_k$ multiplies a term quadratic in the fields,
cf. (\ref{2.6}). The feature that $R_k$ acts as a good infrared
cutoff is not essential for this. In particular, one can easily
write down an exact equation for the dependence of the effective
action on the chemical potential \cite{BJW99chem}. Another interesting
exact equation describes the effect of a variation of the
microscopic mass term for a field, as, for example, the
current quark mass in QCD. In some cases an additional UV-regularization
may be necessary since the UV-finiteness of the momentum
integral in (\ref{2.17}) may not be automatic.
\end{enumerate}

\subsection{Truncations}

Even though intuitively simple, the replacement of the (RG improved)
classical propagator by the full propagator turns the solution of the flow
equation~(\ref{2.17}) into a difficult mathematical problem: The evolution
equation is a functional differential equation. Once $\Gamma_k$ is expanded
in
terms of invariants (e.g.~Eqs.(\ref{2.13}), (\ref{2.14})) this is equivalent
to an infinite system of coupled non--linear partial differential equations.
General methods for the solution of functional
differential equations are not developed very much. They are 
restricted mainly 
to iterative procedures that can be applied once some small expansion
parameter is identified.  They cover usual
perturbation theory in the case of
a small coupling, the $1/N$--expansion or expansions in the dimensionality
$4-d$ or $d-2$. For the flow equation
(\ref{2.17}) they have also been  extended to less
familiar expansions like a
derivative expansion \cite{TW94-1} which is related in critical three-dimensional scalar
theories to a small anomalous dimension \cite{ILF}. 
In the absence of a clearly
identified small parameter one needs to truncate the most
general
form of $\Gamma_k$ in order to reduce the infinite system of coupled
differential equations to a (numerically) manageable size. This truncation
is
crucial. It is at this level that approximations have to be made and, as for
all non-perturbative analytical methods, they are often not easy to control.

The challenge for non-perturbative systems like
critical phenomena in statistical physics or low momentum QCD is to find
flow equations which (a) incorporate all the relevant dynamics so that
neglected effects make only small changes, and (b) remain of manageable
size.
The difficulty with the first task is a reliable estimate of the error. For
the second task the main limitation is a practical restriction for numerical
solutions of differential equations to functions depending only on a small
number of variables.  The existence of an exact functional differential flow
equation is a very useful starting point and guide for this task. At this
point the precise form of the exact flow equation is quite important.
Furthermore, it can be used for systematic expansions through enlargement of
the truncation and for an error estimate in this way.  Nevertheless, this is
not all. Usually, physical insight into a model is necessary in
order to device a
useful non-perturbative truncation!

Several approaches to non-perturbative truncations
have been explored so far:
\begin{itemize}
\item[(i)] {\it Derivative expansion}. We can expand in the number of derivatives  
($\rho\equiv\frac{1}{2}\phi_a\phi^a$)
\begin{equation}
\label{2.29}
  \Gamma_k[\phi]=\int d^d x\left\{
  U_k(\rho)+\frac{1}{2}Z_{k}(\rho)
  \partial_\mu\phi^a\partial^\mu\phi_a+
  \frac{1}{4}Y_{k}(\rho)\partial_\mu\rho
  \partial^\mu\rho+
  \O(\partial^4)\right\} \, .
\end{equation}
The lowest level only includes the scalar potential and a standard
kinetic term. The first correction includes 
the $\rho$-dependent wave function renormalizations 
$Z_{k}(\rho)$ and $Y_{k}(\rho)$. The next level involves then
invariants with four derivatives etc.

One may wonder if a derivative expansion has any chance to account
for the relevant physics of critical phenomena, in a situation where we
know that the critical propagator is non-analytic in the
momentum\footnote{See \cite{Myerson,Golner} for 
early applications of the derivative expansion to critical phenomena.
For a recent study on convergence properties of the derivative
expansion see \cite{MTi99}.}. 
The reason why it can work is that the nonanalyticity builds
up only gradually as $k\to 0$. At the critical temperature a typical
qualitative form of the inverse average propagator is
\beq\label{2.31}
\Gamma_k^{(2)}\sim q^2(q^2+ck^2)^{-\eta/2}\eeq
with $\eta$ the anomalous dimension. Thus the behavior for $q^2\to 0$
is completely regular as long as $k\not=0$.
In addition, the contribution
of fluctuations with small momenta $q^2\ll k^2$ to the flow
equation is suppressed by the
IR cutoff $R_k$. For $q^2\gg k^2$ the ``nonanalyticity'' of the
propagator is already manifest. The contribution of this region
to the momentum integral in (\ref{2.17}) is, however, strongly
suppressed by the derivative $\partial_k R_k$. For cutoff
functions of
the type (\ref{2.7}) only a small momentum range centered around
$q^2\approx k^2$ contributes substantially to the momentum integral
in the flow equation. This suggests the use of a hybrid derivative
expansion where the momentum dependence of $\Gamma_k-\int d^dxU_k$ is
expanded around $q^2=k^2$. Nevertheless, because of the qualitative
behavior (\ref{2.31}), also an expansion around $q^2=0$ should yield
valid results. We will see in section \ref{secsec} that the first
order in the derivative expansion (\ref{2.29}) gives a quite
accurate description of critical phenomena in three-dimensional
$O(N)$ models, except for an (expected) error in the anomalous
dimension. 
\item[(ii)] {\it Expansion in powers of the fields}. As an
alternative ordering principle one may expand $\Gamma_k$
in $n$-point functions $\Gamma_k^{(n)}$ 
\begin{equation}
\label{2.30}
  \Gamma_k[\phi]=
  \sum_{n=0}^\infty\frac{1}{n!}\int
  \left(\prod_{j=0}^n d^d x_j
  \left[\phi(x_j)-\phi_0\right]\right)
  \Gamma_k^{(n)}(x_1,\ldots,x_n)\; .
\end{equation}
If one chooses \cite{Wet91-1}\footnote{See also \cite{TW94-1,Aoki,AMSST96}
for the importance of expanding around $\phi=\phi_0$ instead of
$\phi=0$ and refs.\ \cite{MOP,HKLM,Mor94-2}.}
$\phi_0$ as the $k$--dependent expectation value of $\phi$, 
the series~(\ref{2.30})
starts effectively at $n=2$. 
The flow equations for the 1PI Green
functions $\Gamma_k^{(n)}$ are obtained by functional differentiation
of~(\ref{2.17}).  Similar equations have been discussed first 
in \cite{Wei76-1} from a somewhat different viewpoint. They can also be
interpreted
as a differential form of Schwinger--Dyson equations \cite{DS49-1}.
\item[(iii)] {\it Expansion in the canonical dimension.} We can classify
the couplings according to their canonical dimension. For this 
purpose we expand $\Gamma_k$ around some constant field $\rho_0$
\bea
\Gamma_k[\phi] &=& 
\int d^d x \Big\{ U_k(\rho_0)+U_k^\prime (\rho_0)(\rho-\rho_0)
+\frac{1}{2} U_k^{\prime\prime}(\rho_0)(\rho-\rho_0)^2 + \ldots 
\nonumber\\
&& -\frac{1}{2}\Big(Z_k(\rho_0)+Z_k^\prime (\rho_0)(\rho-\rho_0)
+\frac{1}{2} Z_k^{\prime\prime}(\rho_0)(\rho-\rho_0)^2 + \ldots \Big)
\phi^a \partial_\mu \partial^\mu \phi_a \nonumber\\
&& + \frac{1}{2} \Big(\dot{Z}_k(\rho_0)+\dot{Z}_k^\prime (\rho_0)(\rho-\rho_0)
+ \ldots \Big) \phi^a (\partial_\mu \partial^\mu)^2 \phi_a \nonumber\\
&& - \frac{1}{4} Y_k(\rho_0) \rho \partial_\mu \partial^\mu \rho + \ldots 
\Big\} \, .
\eea
The field $\rho_0$ may depend on $k$. In particular,
for a potential $U_k$ with minimum at $\rho_0(k)>0$
the location of the minimum can be used as one of the 
couplings. In this case $\rho_0(k)$ replaces the 
coupling $U_k^\prime(\rho_0)$ since $U_k^\prime(\rho_0(k))=0$.
In three dimensions one may start by considering an approximation
that takes into account only couplings with positive canonical mass 
dimension, i.e.\ $U^\prime_k(0)$ with mass dimension $M^2$ 
and $U^{\prime\prime}_k(0)$ with dimension $M^1$
in the symmetric regime (potential minimum at $\rho=0$). 
Equivalently, in the spontaneously
broken regime (potential minimum for $\rho\not =0$)
we may take $\rho_0(k)$ and $U^{\prime\prime}_k(\rho_0)$.  
The first correction includes then the dimensionless
parameters $U^{\prime\prime\prime}_k(\rho_0)$ and
$Z_k(\rho_0)$. The second correction includes
$U_k^{(4)}(\rho_0)$, $Z_k^\prime(\rho_0)$ and
$Y_k(\rho_0)$ with mass dimension $M^{-1}$ and so on. 
Already the inclusion of the dimensionless couplings
gives a very satisfactory description of critical
phenomena in three-dimensional scalar theories \cite{TW94-1}.
\end{itemize}

\subsection{Flow equation for the average potential}
\label{FlowAveragePot}

For a discussion of the ground state, its preserved or sponataneously
broken symmetries and the mass spectrum of excitations the
most important quantity is the average potential $U_k(\rho)$.
In the absence of external sources
the minimum $\rho_0$ of $U_{k\to0}$ determines the expectation
value of the order parameter. The symmetric phase with unbroken
$O(N)$ symmetry is realized if $\rho_0(k\to0)=0$, whereas spontaneous
symmetry breaking occurs for $\rho_0(k\to0)>0$. Except for the wave
function renormalization to be discussed later the squared particle
masses $M^2$ are given by $M^2\sim U'(\rho_0=0)$ for
the symmetric phase. Here primes denote derivatives with
respect to $\rho$. For $\rho_0\not=0$ one finds a radial mode
with $M^2\sim U'(\rho_0)+2\rho_0U''(\rho_0)$ and $N-1$
Goldstone modes with $M^2\sim U'(\rho_0)$. For vanishing
external sources the Goldstone modes are massless.

Therefore, we want to concentrate on the flow of $U_k(\rho)$.
The exact flow equation is obtained by evaluating eq. (\ref{2.17}) for a
constant value of $\varphi_a$, say $\varphi_a(x)=\varphi\delta_{a1},
\rho=\frac{1}{2}\varphi^2$. One finds the exact equation 
\cite{Wet91-1,Wet93-2}
\beq\label{2.32}
\partial_tU_k(\rho)=\frac{1}{2}\int\frac{d^dq}{(2\pi)^d}
\frac{\partial R_k}{\partial t}\left(\frac{N-1}{M_0}+
\frac{1}{M_1}\right),
\eeq
with
\bea\label{2.33}
M_0(\rho,q^2)&=&Z_k(\rho,q^2)q^2+R_k(q)+U_{k}'(\rho)
\nonumber\\
M_1(\rho,q^2)&=&\tilde Z_k(\rho,q^2)q^2+R_k(q)+U_{k}'(\rho)
+2\rho U_k''(\rho)\eea
parametrizing the $(a,a)$ and (1,1) element of $\Gamma_k^{(2)}
+R_k\ (a\not=1)$.

As expected, this equation is not closed
since we need information about the $\rho$ and $q^2$-dependent
wave function renormalizations $Z_k$ and $\tilde Z_k$ for
the Goldstone and radial modes, respectively.
The lowest order in the derivative expansion would take
$\tilde Z_k=Z_k=Z_k(\rho_0,k^2)$ independent of $\rho$
and $q^2$, so that only the
anomalous dimension
\beq\label{2.34}
\eta=-\frac{\partial}{\partial t}\ln Z_k\eeq
is needed in addition to the partial differential equation (\ref{2.32}).
For a first discussion let us also neglect the contribution 
$\sim \partial_t Z_k$ in $\partial_t R_k$ and write
\beq\label{A.34}
\frac{\partial}{\partial t}U_k(\rho)=2v_dk^d\left[(N-1)l^d_0\left(
\frac{U_k'(\rho)}{Z_kk^2}\right)+l^d_0\left(\frac{U_k'(\rho)+2\rho
U_k''(\rho)}{Z_kk^2}\right)\right]\eeq
with
\beq\label{AA.34}
v_d^{-1}=2^{d+1}\pi^{d/2}\Gamma\left(\frac{d}{2}\right)\ ,
v_2=\frac{1}{8\pi}\ ,\
v_3=\frac{1}{8\pi^2}\ ,\ v_4=\frac{1}{32\pi^2}.
\eeq
Here we have introduced the dimensionless threshold function
\beq\label{B.34}
l^d_0(w)=\frac{1}{4}v^{-1}_dk^{-d}\int\frac{d^dq}{(2\pi)^d}
\frac{\partial_t(R_k(q)/Z_k)}{q^2+Z^{-1}_kR_k(q)+k^2w}.
\eeq
It depends on the renormalized particle mass $w=M^2/(Z_kk^2)$ and
has the important property that it decays rapidly for $w\gg1$.
This describes the decoupling of modes with renormalized squared mass
$M^2/Z_k$ larger than $k^2$. In consequence, only modes with mass
smaller than $k$ contribute to the flow. The flow equations ensure
automatically the emergence of effective theories for the low-mass modes! 
The explicit form of the threshold functions (\ref{B.34})
depends on the choice of $R_k$.  For a given explicit form
of the threshold functions eq.\ (\ref{A.34}) turns into a 
nonlinear partial differential
equation for a function $U$ depending on the two variables
$k$ and $\rho$. This can be solved numerically by appropriate
algorithms \cite{ABBFTW95-1}.   

Eq.\ (\ref{A.34}) was first derived
\cite{Wet91-1} as a renormalization group improved 
perturbative expression and its intuitive form close to perturbation theory
makes it very suitable for practical investigations. 
Here it is important to note that the use of the average action 
allows for the inclusion of propagator corrections
(wave function renormalization effects) in a direct and
systematic way.
Extensions to more complicated scalar models
or models with fermions \cite{Wet90-1} are straightforward.
In the limit of a sharp cutoff 
and for vanishing anomalous dimension eq.\ (\ref{A.34})
coincides with the Wegner-Houghton equation \cite{WH73-1} for the
potential, first discussed in \cite{NCS1} (see also
\cite{NC2,Has86-1,CT97}). 

\begin{figure}[h]
\unitlength1.0cm
\begin{center}
\begin{picture}(13.,9.)
\put(6.4,-0.2){\footnotesize $\varphi$ }
\put(1.75,8.){$U_k(\varphi)$}

\put(-0.5,0.){
\epsfysize=13.cm
\epsfxsize=9.cm
\rotate[r]{\epsffile{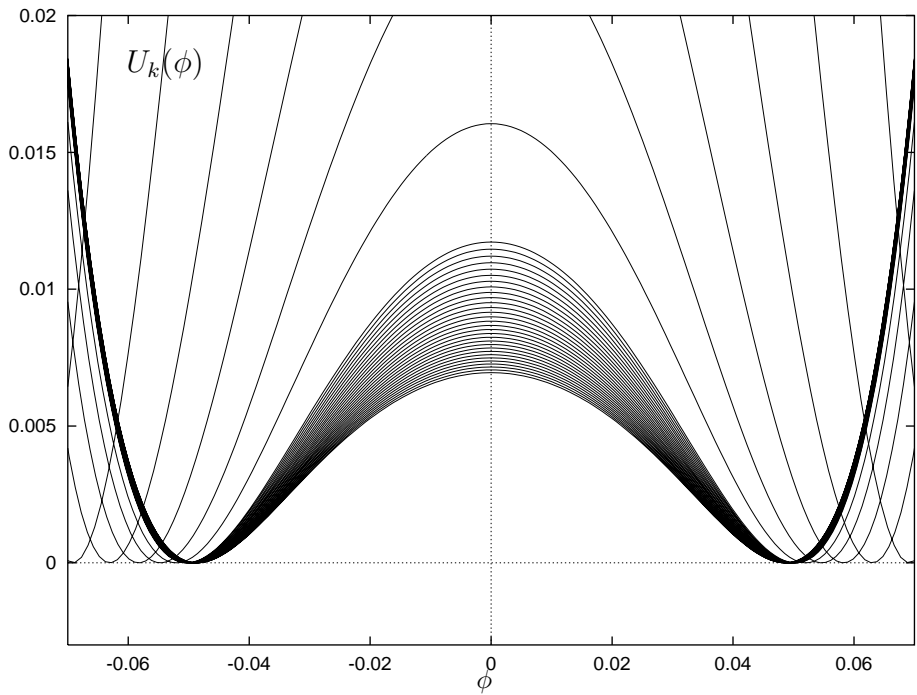}}
}
\end{picture}
\end{center}
\caption[]{
\em Average potential $U_k(\varphi)$ for different scales $k=e^t$.
The shape of $U_k$ is displayed 
 at smaller intervals $\Delta t=-0.02$ after the minimum has settled at
 a constant value.
 This demonstrates the approach to convexity in the ``inner region'', while
the ``outer region'' becomes $k$-independent for $k \to 0$.  }
\label{apptoconvexity}
\end{figure}

Eq.\ (\ref{A.34}) can be used as a practical
starting point for various systematic expansions.
For example, it is the lowest order in the derivative 
expansion. The next order includes $q^2$-independent
wave function renormalizations $Z_k(\rho), \tilde Z_k(\rho)$
in eq. (\ref{2.33}). For $N=1$ the first-order in the
derivative expansion leads therefore to coupled partial
nonlinear differential equations for two functions $U_k(\rho)$ and
$Z_k(\rho)$ depending on two variables $k$ and $\rho$.
We have solved these differential equations numerically
and the result for different values of $k$
is plotted in fig.\ \ref{apptoconvexity}. 
The initial values of 
the integration correspond to the phase with spontaneous symmetry
breaking. More details can be found in sect.\ \ref{secsec}. 

\section{$O(N)$-symmetric scalar models}
\label{secsec}

In this section we study the $N$-component scalar
model with $O(N)$-symmetry in three and two dimensions.
Due to ``triviality'' the four-dimensional potential flows
rapidly to the perturbative domain \cite{pw} and needs not to be
discussed in detail here.
The $O(N)$ model serves as a prototype for investigations concerning 
the restoration of a spontaneously broken symmetry at 
high temperature. For $N=4$ the model describes the scalar
sector of the electroweak standard model in the limit of
vanishing gauge and Yukawa couplings. It is
also used as an effective model for the chiral
phase transition in QCD in the limit of two quark flavors. 
 In condensed matter physics $N=3$ corresponds to the
Heisenberg model used to describe the ferromagnetic
phase transition. There are other applications like the
helium superfluid transition ($N=2$), liquid-vapor transition
($N=1$) or statistical properties of long polymer chains
($N=0$). 

\subsection{Scaling form of the exact flow equation for the potential}
\label{ScalingFormPot}

Let us first  derive an explicitly scale-invariant form of the exact flow equation (\ref{2.32}).
This will be a useful starting point for the discussion
of critical phenomena. It is convenient
to use a dimensionless cutoff function
\beq\label{2.35}
r_k\left(\frac{x}{k^2}\right)=\frac{R_k(x)}{Z_kx}\ ,\
x\equiv q^2\ ,\ Z_k=Z_k(\rho_0,0)\eeq
and write the flow equations as
\beq\label{2.36}
\frac{\partial}{\partial t}U_k(\rho)=v_d\int^\infty_0dx\ x^{\frac{d}{2}}
s_k\left(\frac{x}{k^2}\right)\left(\frac{N-1}
{M_0/Z_k}+\frac{1}{M_1/Z_k}\right)\eeq
with
\beq\label{2.38}
s_k\left(\frac{x}{k^2}\right)=\frac{\partial}{\partial t}r_k
\left(\frac{x}{k^2}\right)-\eta r_k\left(\frac{x}{k^2}\right)=
-2x\frac{\partial}{\partial x}r_k\left(\frac{x}{k^2}\right)-\eta 
r_k\left(\frac{x}{k^2}\right).\eeq
We parametrize the wave function renormalization by
\beq\label{2.39}
z_k(\rho)=\frac{Z_k(\rho,0)}{Z_k}\ ,\ z_k(\rho_0)\equiv1\ ,\
\rho \tilde y_k(\rho)=
\frac{\tilde Z_k(\rho,0)-Z_k(\rho,0)}{Z_0}k^{d-2},\eeq
so that
\beq\label{2.40}
\frac{\partial}{\partial t}U_k=2v_dk^d[(N-1)l^d_0\left(
\frac{U_k'}{Z_kk^2};\eta,z_k\right)+l^d_0\left(
\frac{U_k'+2\rho U_k''}{Z_kk^2}; \eta,z_k+
Z_k\rho \tilde y_kk^{2-d}\right)]+\
\Delta\zeta_kk^d,\eeq
where $l_0^d(w;\eta, z)$ is a
generalized dimensionless threshold function $(y=x/k^2)$
\beq\label{2.41}
l^d_0(w;\eta,z)=\frac{1}{2}\int^\infty_0dy y^{\frac{d}{2}}s_k(y)
[(z+r_k(y))y+w]^{-1}.\eeq
The correction
$\Delta \zeta_k$ contributes only in next (second) order in a derivative expansion and will be omitted.
Finally, we may remove the explicit dependence on $Z_k$ and $k$
by using scaling variables
\beq\label{2.43}
u_k=U_kk^{-d},\quad \tilde \rho=Z_kk^{2-d}\rho\ ,\quad
\kappa=Z_kk^{2-d}\rho_0\eeq
Evaluating the $t$-derivative at fixed $\tilde\rho$ and
using the notation $u'=\partial u/\partial\tilde\rho$ etc. one
obtains the scaling form of the exact evolution equation
for the average potential
\beq\label{2.44}
\partial_tu_{|_{\tilde\rho}}=-du+(d-2+\eta)\tilde\rho u'
+2v_d\left\{(N-1)l^d_0(u';\eta,z)
+l^d_0(u'+2\tilde\rho u'',\eta,z+\tilde\rho \tilde y)\right\} 
\eeq
All explicit dependence on the scale $k$ or the wave function
renormalization $Z_k$ has disappeared.
Reparametrization invariance
under field scaling is obvious in this form. For $\rho\to\alpha^2
\rho$ one also has $Z_k\to \alpha^{-2}Z_k$ so that
 $\tilde\rho$ is invariant. This property needs the factor $Z_k$
in $R_k$. This version is therefore most
appropriate for a discussion of critical behavior.
The universal features of the critical behavior for second-order
phase transitions are related to the existence of a
scaling solution\footnote{More
precisely, $\partial_t u'=0$ is a sufficient condition for the
existence of a scaling solution. 
The universal aspects of first-order transitions are
also connected to exact or approximate scaling solutions.}.  
This scaling solution solves the differential
equation for a $k$-independent function $u(\tilde\rho)$,
which results from (\ref{2.44}) by setting 
$\partial_tu=0$, and similar for
$z(\tilde\rho)$ etc..
In lowest order in the derivative expansion with
a constant wave function renormalization the scaling
potential can be directly obtained by solving the second-order differential 
equation $\partial_tu=0$. 
It has been shown that, of all possible solutions, the physical fixed point
corresponds to the solution $u(\tilde{\rho})$ which is non-singular 
in the field \cite{Felder,FB,Mor94-2,Mor94-1}. For $d=3$ the 
only nontrivial solution corresponds to the Wilson-Fisher
\cite{WilFis72} fixed point.

In first nontrivial order in the derivative expansion one has to
supplement two 
partial differential equations \cite{Wet91-1,Wet93-2,TW94-1}
for the $\tilde\rho$-dependent wave 
function renormalizations $z$ and $\tilde z=z+\tilde\rho\tilde y$
\begin{eqnarray}
\partial_t z
	&=&\eta z +\tilde\rho z'(d-2+\eta)\nonumber\\
	&&-(4v_d/d)\tilde\rho^{-1}\bigl\{\bigr.
	m^d_{2,0}(w,z,\eta)-2m^d_{1,1}(w,\tilde w,z,\tilde z,\eta)
	+m^d_{0,2}(\tilde w,\tilde z,\eta)\bigl.\bigr\}\nonumber\\
	&&-2v_d(\tilde z-z)\tilde\rho^{-1}\bigl\{\bigr.l_1^d(\tilde w,\tilde z,\eta)-(2/d)(\tilde z-z)l^{d+2}_2(\tilde w,\tilde z,\eta)\bigl.\bigr\}\nonumber\\
	&&-2v_dz'\bigl\{\bigr.(N-1)l_1^d(w,z,\eta)
	-(8/d)n^d_{1,1}(w,\tilde w,z,\tilde z,\eta)\nonumber\\
	&& +\left(5+2z''\tilde\rho/z'\right)
	l_1^d(\tilde w,\tilde z,\eta)
	-(4/d)z'\tilde\rho l_{1,1}^{d+2}(w,\tilde w,z,\tilde z,\eta)\bigl.\bigr\},
\label{dzdt}
\end{eqnarray}

\begin{eqnarray}
\partial_t\tilde z
	&=&\eta\tilde z +\tilde\rho\tilde z'(d-2+\eta)\nonumber\\
&&	-2v_d(\tilde z'+2\tilde\rho\tilde z'')l_1^d(\tilde w, \tilde z,\eta)
	+8v_d\tilde\rho\tilde z'(3u''+2\tilde\rho u''')
	l_2^d(\tilde w, \tilde z,\eta)\nonumber\\
&&	+4v_d\left(2+1/d\right)\tilde\rho(\tilde z')^2
	l_2^{d+2}(\tilde w, \tilde z,\eta)
	-(8/d)v_d\tilde\rho(3u''+2\tilde\rho u''')^2
	\tilde m_4^d(\tilde w, \tilde z,\eta)\nonumber\\
&&	-(16/d)v_d\tilde\rho\tilde z'(3u''+2\tilde\rho u''')
	\tilde m_4^{d+2}(\tilde w, \tilde z,\eta)
	-(8/d)v_d\tilde\rho(\tilde z')^2
	\tilde m_4^{d+4}(\tilde w, \tilde z,\eta)\nonumber\\
&&	+(N-1)v_d\bigl\{\bigr.
	-2\left(\tilde z'-\tilde\rho^{-1}(\tilde z -z)\right)
	l_1^d(w,z,\eta)
	-(8/d)\tilde\rho (u'')^2m_4^d(w,z,\eta)\nonumber\\
&&	-(16/d)\tilde\rho u''z'm_4^{d+2}(w,z,\eta)
	-(8/d)\tilde\rho (z')^2m_4^{d+4}(w,z,\eta)
	+4(\tilde z -z)u''l_2^d(w,z,\eta)\nonumber\\
	&&\bigl .+4\left(z'(\tilde z -z)+
	(1/d)\tilde\rho(z')^2\right)
	l_2^{d+2}(w,z,\eta)\bigr \}	
\label{dztdt}
\end{eqnarray}
They invoke the ``threshold functions'' $t^d_{n_1n_2}=\{
l_{n_1,n_2}^d$, $m_{n_1,n_2}^d$, $\tilde m_{n_1,n_2}^d, n_{n_1,n_2}^d\}$ defined by the integrals
\begin{equation}
t^d_{n_1,n_2}=-\frac{1}{2}\intop_0^\infty 
	dyy^{\frac{d}{2}-1}
	\tilde{\partial_t}\left \{ \frac{X}{(p(y)+w)^{n_1}(\tilde p(y)+\tilde w)^{n_2}}\right \},
\end{equation}
with $X=1,\ y (\partial_yp)^2,\ y(\partial_y\tilde p)^2,\ y\partial_y p$ for $l,\ m,\ \tilde m,\ n$ respectively.
We have defined $w=u',\ \tilde w=u'+2\tilde\rho u'',\ 
\tilde z=z+\tilde \rho\tilde y,\ 
p(y)=y(z+r(y))$, and  $\tilde p(y)=y(\tilde z+r(y)).$
The derivative $\tilde \partial_t$ only acts on the $k$-dependence of the cutoff $R_k$, i.e. 
$\tilde\partial_t p(y)=-y\bigl(\eta r(y)+2y\partial_yr(y)\bigr)$
and we note that $\tilde\partial_t\partial_yp=\partial_y\tilde\partial_tp$. Finally we abbreviate $l_{n,0}^d=l_n^d$ etc., and $l_0^d$ (\ref{2.41})
corresponds to the rule
$(p+w)^{-n}\rightarrow-\log(p+w)$.
The expression for the anomalous dimension $\eta=-\partial_t\ln Z_k$ can be obtained from the identity $\partial_t z_k(\kappa_k)\equiv 0$, 
which corresponds to a definition
of $Z_k$ by $z_k(\kappa)=1$. 
(For $N=1$ one has $\tilde y=0$, $z=\tilde z$). 

These equations are valid in arbitrary dimension. 
We will show that they lead to quantitatively accurate results.
This is done by a numerical solution with initial values specified at a microscopic scale $k=\Lambda$.

\subsection{Threshold functions}
\label{SecThresh}

In situations where the momentum dependence of the propagator can
be approximated by a standard form of the kinetic term and is
weak for the other 1PI correlation functions, the ``non-perturbative''
effects beyond one loop arise to a large extent from the threshold
functions. We will therefore discuss their most
important properties and introduce the notation
\bea\label{2.45}
&&l^d_n(w;\eta,z)=\frac{n+\delta_{n,0}}{4}v_d^{-1}k^{2n-d}\int
\frac{d^dq}{(2\pi)^d}\partial_tR_k(q)(Z_kzq^2+R_k(q)+Z_kwk^2)^{-(n+1)},
\nonumber\\
&&l^d_{n+1}(w;\eta,z)=-\frac{1}{n+\delta_{n,0}}\frac{\partial}
{\partial w}l^d_n(w;\eta,z),\nonumber\\
&&l^d_n(w;\eta)=l^d_n(w;\eta,1)\ ,\ l^d_n(w)=l^d_n(w;0,1)\ ,\ l^d
_n=l^d_n(0).\eea
The precise form of the threshold functions depends on the
choice of the cutoff function $R_k(q)$. There are, however,
a few general features which are independent of the particular scheme:
\begin{enumerate}
\item
For $n=d/2$ one has the universal property
\beq\label{2.46}
l^{2n}_n=1.\eeq
This guarantees the universality of the perturbative
$\beta$-functions for the quartic coupling in $d=4$ or for the coupling
in the nonlinear $\sigma$-model in $d=2$.

\item
If the momentum integrals are dominated by $q^2\approx k^2$
and $R_k(q)\lta k^2$, the
threshold functions obey for large $w$
\beq\label{2.47}
l^d_n(w)\sim w^{-(n+1)}.\eeq

\item
The threshold functions diverge for some negative value
of $w$. This is related to the fact that the average potential
must become convex for $k\to0$.
\end{enumerate}

It is instructive to evaluate the threshold functions 
$l^d_n(w)$ explicitly for
a simple cutoff function of the form
\beq\label{C.34}
R_k=Z_k(k^2-q^2)\Theta(k^2-q^2)\eeq
where $(x=q^2)$
\beq\label{D.34}
l^d_0(w)=k^{2-d}\int^{k^2}_0dxx^{\frac{d}{2}-1}
[(w+1)k^2]^{-1}
+k^{2-d}\int^\infty_0dxx^{\frac{d}{2}-1}\frac{\delta(x-k^2)(k^2-x)}
{x+(k^2-x)\Theta(k^2-x)+k^2w}.
\eeq
This corresponds \cite{LP} to an ``optimal cutoff'' in the sense of
\cite{LitimS}.
The second term in the expression for $l^d_0$ has to be properly defined
and we consider eq. (\ref{C.34}) as the limit $\gamma\to\infty$ of a family
of cutoff functions, e.g.
\beq\label{E.34}
R_k=\frac{2\gamma}{1+\gamma}Z_k\left(k^2-(1+\frac{2}{\gamma^2+1}
-\frac{6}{\gamma^2+2})q^2\right)\left(\frac{q^2}{k^2}\right)^\gamma
\left[\exp\left\{\frac{2\gamma}{1+\gamma}\left(\frac{q^2}{k^2}\right)
^\gamma\right\}-1\right]^{-1}.\eeq
This confirms that the second term in eq. (\ref{D.34})
vanishes and one obtains the threshold functions
\beq\label{F.34}
l^d_0(w)=\frac{2}{d(1+w)}\ ,\quad l^d_n(w)=
\frac{2n}{d(1+w)^{n+1}}\quad {\rm for}\quad n\geq 1\eeq
 
In leading order in the derivative expansion $(z=1, \tilde y
=0)$ and neglecting the anomalous dimension in $l^d_0$,
this choice of the threshold function yields the simple evolution equation
\beq\label{2.52}
\partial_tu=-du+(d-2+\eta)\tilde\rho \
u'+\frac{4v_d}{d}\left\{\frac{N-1}{1+u'}+\frac
{1}{1+u'+2\tilde\rho u''}\right\}.\eeq
In particular, for $N=1, d=3$ and neglecting the anomalous dimension
we end with a simple partial differential equation
\beq\label{3.24}
\partial_tu=-3u+\tilde\rho u'+\frac{1}{6\pi^2(1+u'+2\tilde\rho u'')}\eeq
For our numerical calculations we mainly use
the smooth exponential cutoff (\ref{2.7}) which is qualitatively
similar to (\ref{C.34}) and corresponds to (\ref{E.34}) with 
$\gamma=1$. 

\subsection{Second-order phase transitions}
\label{SecTrans}

At a second-order phase transition 
there is no mass scale present in the theory.
In particular, one 
expects a scaling behavior of the rescaled effective 
average potential $u_k(\tilde{\rho})$. Let us follow 
a typical trajectory which describes the scale dependence
of the derivative $u_k'(\tilde{\rho})$ as $k$ is lowered from $\Lambda$ to zero.  
Near the phase transition the   
trajectory spends most of the ``time'' $t$ in the vicinity of
the $k$-independent scaling solution given by
$\partial_t u'_* (\tilde{\rho}) = 0$.
Only at the end of the running the ``near-critical''
trajectories deviate from the scaling solution. For
$k \rightarrow 0$ they either end up in the symmetric phase with
$\kappa =0$ and positive constant mass term $m^2$ so that
$u'_k(0) \sim m^2/k^2$; or they lead to a non-vanishing
constant $\rho_0$ indicating spontaneous symmetry breaking with
$\kappa \rightarrow Z_0 k^{2-d} \rho_0$. The equation of state
involves the potential $U_0(\rho)$ for
temperatures away from the critical temperature.
Its computation requires the
solution for the running 
away from the critical trajectory.

More precisely, we start at short distances
$(k=\Lambda)$ with a quartic potential or
\beq
u'_{\Lambda}(\tilde{\rho}) = \lambda_\Lambda (\tilde{\rho} - \kappa_\Lambda).
\label{eighteen} \eeq
We arbitrarily choose $\lambda_\Lambda=0.1$ and fine tune $\kappa_\Lambda$
so that a scaling solution is approached at the later stages
of the evolution. There is a critical value
$\kappa_{cr}$
for which the evolution leads to the scaling solution.
For the results in fig. 1 
a value $\kappa_\Lambda$ slightly larger than $\kappa_{cr}$
is used. As $k$ is lowered
(and $t$ turns negative), $u'_k(\tilde{\rho})$ deviates from its initial
linear shape. Subsequently it evolves towards a form which is
independent of $k$ and corresponds to the scaling
solution $\partial_t u'_* (\tilde{\rho}) = 0$. It spends a long ``time''
$t$ -- which can be rendered arbitrarily long through appropriate
fine tuning of $\kappa_\Lambda$ --
in the vicinity of the scaling solution. During this
``time'', the minimum of the potential $u'_k(\tilde{\rho})$
takes a fixed value $\kappa_*$,
while the minimum of $U_k(\rho)$ evolves towards zero according to
\beq
\rho_0(k) = k \kappa_* / Z_k.
\label{nineteen} \eeq
The longer $u'_k(\tilde{\rho})$ stays near the scaling solution, the
smaller the resulting value of $\rho_0(k)$ when the
system deviates from it.
As this value determines the mass scale for the
renormalized theory at $k=0$, the
scaling solution governs the behavior of the system
very close to the phase transition, where the characteristic
mass scale goes to zero.
Another important property of the ``near-critical''
trajectories, which spend a long ``time'' $t$ near
the scaling solution, is that they become insensitive
to the details of the short-distance interactions that determine the
initial conditions for the evolution. In particular,
after $u'_k(\tilde{\rho})$ has
evolved away from its scaling form $u'_*(\tilde{\rho})$, its
shape is independent of the choice of $\lambda_\Lambda$ for
the classical theory.
This property gives rise to the universal behavior
near second-order phase transitions.
For the solution depicted in
fig.\ 1, 
$\kappa$ finally grows in such a way that
$\rho_0(k)$ approaches a constant value
for $k \rightarrow 0$.

As we have already mentioned the details of
the renormalized theory in the vicinity of the phase
transition are independent of the classical coupling
$\lambda_\Lambda$. Also the
initial form of the potential does not have
to be of the quartic form of eq.\ (\ref{eighteen})
as long as the symmetries are respected.
The critical theory can
be parameterized in terms of critical exponents
\cite{WilFis72}. An example 
is the anomalous dimension $\eta$, which determines the
behavior of the two-point function at the critical
temperature according to eq. (\ref{2.31}) for $k\to 0$.
Its value is given by the value of
$\eta$ for the scaling solution.
The critical exponents are universal quantities that depend
only on the dimensionality of the system and its internal
symmetries. For our three-dimensional theory they depend only on the
value of $N$ and
can be easily extracted from our results. We concentrate here on
the exponent $\nu$ which parameterizes the behavior of the
renormalized mass in the critical region and consider the symmetric phase
\beq
m^2 =  \frac{1}{Z_k} \frac{d U_k(0)}{d\rho}
= k^2 u'_k(0)~~~~~~~{\rm for}~~k \rightarrow 0.
\label{twentyone} \eeq
The behavior of $m^2$ in the critical region
depends only on the distance from the phase transition, which
can be expressed in terms of the difference of
$\kappa_\Lambda$ from the critical value $\kappa_{cr}$ for which
the renormalized theory has exactly $m^2 =0$.
The exponent $\nu$ is determined from the relation
\beq
m^2 \sim |\delta\kappa_{\Lambda}|^{2 \nu} = 
|\kappa_\Lambda - \kappa_{cr}|^{2 \nu}.
\label{twentytwo} \eeq
Assuming proportionality $\delta\kappa_\Lambda\sim T_c-T$, 
this yields the critical temperature dependence of the 
correlation length $\xi=m^{-1}$.
For a determination of $\nu$ from our results we calculate
$m^2$ for various values of $\kappa_\Lambda$ near $\kappa_{cr}$.
We subsequently plot $\ln(m^2)$ as a function
of $\ln|\delta\kappa_{\Lambda}|$. This curve becomes linear for
$\delta\kappa_{\Lambda} \rightarrow 0$ and we
obtain $\nu$ from the constant slope.
\bigskip

\begin{center}
\begin{tabular}{|c||c|c||c|c|}
\hline
$N$
&\multicolumn{2}{c||}{$\nu$}
&\multicolumn{2}{c|}{$\eta$}
\\

\hline \hline
&
&$0.5882(11)^a$
&
&$0.0284(25)^a$
\\
0
&$0.589^f$
&$0.5875(25)^{b}$
&$0.040^f$
&$0.0300(50)^{b}$
\\

&$0.590^g$
&$0.5878(6)^{c1}$
&$0.039^g$
&
\\

&
&$0.5877(6)^e$
&
&
\\ \hline
&
&$0.6304(13)^a$
&
&$0.0335(25)^{a}$
\\
1
&$0.643^f$
&$0.6290(25)^{b}$
&$0.044^f$
&$0.0360(50)^{b}$
\\

&$0.6307^{g}$
&$0.63002(23)^{c2}$
&$0.0467^{g}$
&$0.0364(4)^{c2}$
\\

&
&$0.6294(9)^e$
&
&$0.0374(14)^e$
\\ \hline

&
&$0.6703(15)^{a}$
&
&$0.0354(25)^{a}$
\\
2
&$0.697^f$
&$0.6680(35)^{b}$
&$0.042^f$
&$0.0380(50)^{b}$
\\

&$0.666^g$
&$0.67166(55)^{c2}$
&$0.049^g$
&$0.0381(3)^{c2}$
\\

&
&$0.6721(13)^e$
&
&$0.042(2)^e$
\\ \hline

&
&$0.7073(35)^{a}$
&
&$0.0355(25)^{a}$
\\
3
&$0.747^f$
&$0.7045(55)^{b}$
&$0.038^f$
&$0.0375(45)^{b}$
\\

&$0.704^g$
&$0.716(2)^{c1}$
&$0.049^g$
&
\\

&
&$0.7128(14)^e$
&
&$0.041(2)^e$
\\ \hline

&
&$0.741(6)^{a}$
&
&$0.0350(45)^{a}$
\\
4
&$0.787^f$
&$0.737(8)^{b}$
&$0.034^f$
&$0.036(4)^{b}$
\\

&$0.739^g$
&$0.759(3)^{c1}$
&$0.047^g$
&
\\

&
&$0.7525(10)^{e}$
&
&$0.038(1)^{e}$
\\  \hline
10
&$0.904^f$
&$0.894(4)^{c1}$
&$0.019^f$
&
\\

&$0.881^g$
&$0.877^{d}$
&$0.028^g$
&$0.025^{d}$
\\ \hline
100
&$0.990^f$
&$0.989^{d}$
&$0.002^f$
&$0.003^{d}$
\\ 

&
$0.990^g$
&
&$0.003^g$
&
\\
\hline
\end{tabular}
\end{center}

\medskip
\noindent Table 1 {\em
Critical exponents $\nu$ and $\eta$
for various values of $N$.
For comparison we list results obtained with other methods as summarized in
\cite{GZ}, \cite{journal}, \cite{BC97-1}, \cite{CPRV},
\cite{Zinn99}: \\
a) From perturbation series at fixed dimension including seven--loop 
contributions. \\
b) From the $\epsilon$-expansion at order $\epsilon^5$. \\
c) From high temperature expansions (c1: \cite{BC97-1}, c2: \cite{CPRV}, 
see also \protect\cite{Reisz95,ZinnLaiFisher96}). \\
d) From the $1/N$-expansion at order $1/N^2$. \\
e) From lattice Monte Carlo simulations \cite{LMC0,LMC10} (see also 
\cite{hasenbusch,engels}).\\
f) Average action in lowest order in the derivative expansion
\cite{report,TW94-1}.\\
g) From first order in the derivative expansion for the average action
with field dependent wave function renormalizations (see \cite{B} 
for $N=1$ and \cite{vGW} for $N>1$)}.

\bigskip
In table 1
we compare our values for the critical exponents $\nu$ and $\eta$
obtained from the numerical solution of the partial differential
equation (\ref{2.44})
with results obtained from other methods
(such as the $\epsilon$-expansion, perturbation series
at fixed dimension, lattice high temperature expansions, 
Monte Carlo simulations and the $1/N$-expansion).
We show both the lowest order of the derivative expansion
({\it f}) -- which needs an additional equation for
$\eta$ \cite{TW94-1}  -- and the first order ({\it g})
which corresponds to solutions of the 
system of eqs. (\ref{2.44}), (\ref{dzdt}), (\ref{dztdt}).
As expected $\eta$ is rather poorly determined since it is
the quantity most seriously affected by the omission of the
higher derivative terms in the average action. The exponent
$\nu$ is in agreement with the known results at the
1 \% level, with a discrepancy between
lowest and first order $((f)$ vs. $(g))$ roughly equal to the
value of $\eta$ for various $N$. Similar results 
are found for a variety of forms 
for the infrared cutoff function 
\cite{AL,BHLM95-1,AMSST96,CT97,MT97-1}. We observe a convincing
apparent convergence of the derivative expansion towards the
high precision values obtained from the other methods.

\subsection{Kosterlitz-Thouless transition and essential scaling}
\label{SecKost}

In two dimensions $(d=2)$ and for $N=2$ the
Kosterlitz-Thouless phase transition \cite{KT73-1} may describe the critical behaviour of various two dimensional systems. 
It poses a challenge to our theoretical understanding due to several uncommon features. 
The low temperature phase exhibits a massless Goldstone boson like excitation despite the fact that the global $U(1)$ symmetry is not spontaneously broken by a standard order parameter \cite{MW66-1}. 
In this phase the critical exponents depend on the temperature. 
In the high temperature phase the approach to the transition is not governed by critical exponents but rather by essential scaling. These features
can be explained by a description of the phase transition as 
a condensation of vortices \cite{KT73-1}. Within our 
renormalization group equation there is no direct need for the
introduction of vortex degrees of freedom since all nonperturbative
effects are included in the exact equation. We therefore aim here
at a complementary description in terms of the linear $O(N)$ scalar
model without ever introducing vortices. This constitutes an
excellent test for the nonperturbative content of our equation.

Already a quartic polynomial truncation for the potential
shows the characteristic properties of the low temperature phase \cite{GW95-1}. 
This section is based on \cite{vGW}  and 
uses the first order in the derivative expansion (\ref{2.44}), 
(\ref{dzdt}),(\ref{dztdt}). One finds, indeed, essential scaling in the
high temperature phase. Our discussion of the low temperature phase
is expected to hold at the 10-20 percent accuracy level.

In this section we present a unified picture for $O(N)$ models
in $d=2$. The most important parameter is the location $\kappa$
of the potential minimum and we investigate its dependence on $k$.
For $\kappa\gg1$ the evolution is dominated by the $N-1$ Goldstone modes ($N>1$). More precisely, the threshold functions at the minimum vanish rapidly for $\tilde w=2\kappa u''(\kappa)\gg1$. For $d=2$ and 
$N>2$ the effective
coupling of the nonlinear $\sigma$- model for the Goldstone bosons is given by $\kappa^{-1}$ \cite{Wet91-1}. 
The  universality of the $\beta$-function for the nonlinear
 coupling \cite{Pol75-1} implies for $\kappa\gg1$ an asymptotic
 form ($N\geq2, d=2$) 
\begin{equation}
\partial_t\kappa=\beta_\kappa=\frac{N-2}{4\pi}+\frac{N-2}{16\pi^2}\kappa^{-1}+{\cal O}(\kappa^{-2}). \label{betakappa0}
\end{equation}
In the linear description one can easily obtain an equation for $\kappa$ by using $du_k'(\kappa_k)/dt=
\partial_tu'_k(\kappa)+u_k''(\kappa)\partial_t\kappa=0$
together with equation (\ref{2.44}). By evaluating the above equations for large $\kappa$ it is possible to compare with (\ref{betakappa0}). 
In the much simpler 
quartic truncation one already obtains \cite{Wet91-1}
the correct lowest order in the above expansion.
In order to reproduce the exact two loop result one has, however, to go even beyond the first order in the derivative expansion \cite{pw}.  

From equation (\ref{betakappa0}) we expect that $\kappa$ will run only marginally at large $\kappa$. As a consequence the flow of the action follows a single trajectory for large $-t$ and can be characterized by a single scale.      	
Notice that the perturbative $\beta$-function vanishes for $N=2$ since the Goldstone bosons are not interacting in the abelian case. Thus for large $\kappa$ one expects a line of fixed points which can be parameterized by $\kappa$. This fact plays a major role in the discussion of the Kosterlitz-Thouless transition below. It is responsible for the temperature dependence of the critical exponents.
We have evaluated $\beta_\kappa=\beta_\kappa^{(1)}(N-2)/(4\pi)+\beta_\kappa^{(2)}
\kappa^{-1}(N-2)/(16\pi^2)+...$ numerically from the solution of eqs. (\ref{2.44}) - (\ref{dztdt})
and extracted the expansion coefficients for large $\kappa$ (see table 2).
Our numerical result is close to the two-loop
expression.

\bigskip
\begin{center}
\begin{tabular}{|c|lll|ll|}
\hline
N	& $C_{ERGE}$ & $C_{\overline{MS}}$& $C_s$&$\beta_\kappa^{(1)}$&$\beta_\kappa^{(2)}$ \\
\hline
3	&2.81$\pm$0.30	&2.94	&1.00	&1.00	&0.79\\
9	&1.22$\pm$0.03	&1.25	&1.05	&1.00	&0.84\\
100	&1.08$\pm$0.04	&1.02	&1.06	&1.00	&0.87\\
\hline
\end{tabular}
\end{center}

\medskip\noindent Table 2\ 
{\em Nonabelian nonlinear sigma model in $d=2$. We show the ratio between the renormalized mass $m_R$ and the nonperturbative scale $\Lambda_{ERGE}$ in comparison with the known ratio \cite{HasMagNied90} 
invoking $\Lambda_{\overline{MS}}$: $C_{ERGE}=m_R/\Lambda_{ERGE}$, $C_{\overline{MS}}=m_R/\Lambda_{\overline{MS}}$, $C_s=m_R/k_s$. We also display the expansion coefficients for the beta function.}

\bigskip

\begin{figure}[h]
\centering\epsfig{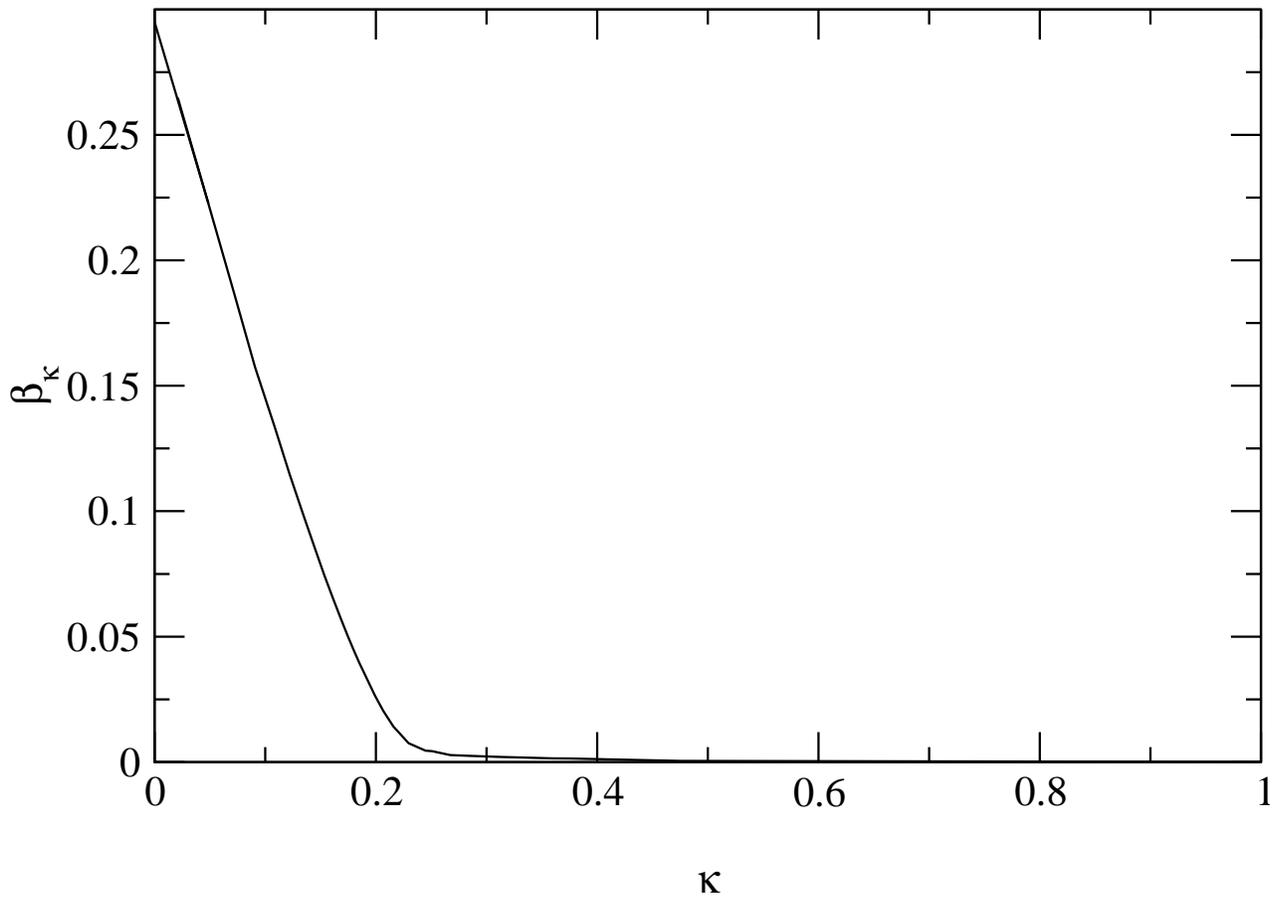}
\caption{The beta function for $d=2,\ N=2$.}
\label{n2betakappa}
\end{figure}

For the nonabelian nonlinear sigma model in $d=2$, $N>2$ there exists an exact expression \cite{HasMagNied90} for the ratio of the renormalized mass $m_R$ and the scale $\Lambda_{\overline{MS}}$ which characterizes the two loop running coupling in the $\overline{MS}$ scheme by dimensional transmutation.
 The flow equation (\ref{2.17}) together with a choice of the cutoff $R_k$ and the initial conditions also defines a renormalization scheme. The corresponding parameter $\Lambda_{ERGE}$ specifies the two loop perturbative value of the running coupling $\kappa^{-1}$ similar to $\Lambda_{\overline{MS}}$ in the $\overline{MS}$ scheme. The numerical solution of the flow equation permits us to compute $m_R/\Lambda_{ERGE}$. (Two loop accuracy would be needed for a quantitative determination of $\Lambda_{ERGE}/\Lambda_{\overline{MS}}$.) In table 2
we compare our results with the exact value of $m_R/\Lambda_{\overline{MS}}$. The quantitative agreement between 
$C_{ERGE}$ and $C_{\overline{MS}}$ is striking and
suggests $\Lambda_{ERGE}\approx \Lambda_{\overline{MS}}$. 
We also report the ratio $m_R/k_s$ with $k_s$ defined by $\kappa(k_s)=0$.

\begin{figure}[h]
\centering\epsfig{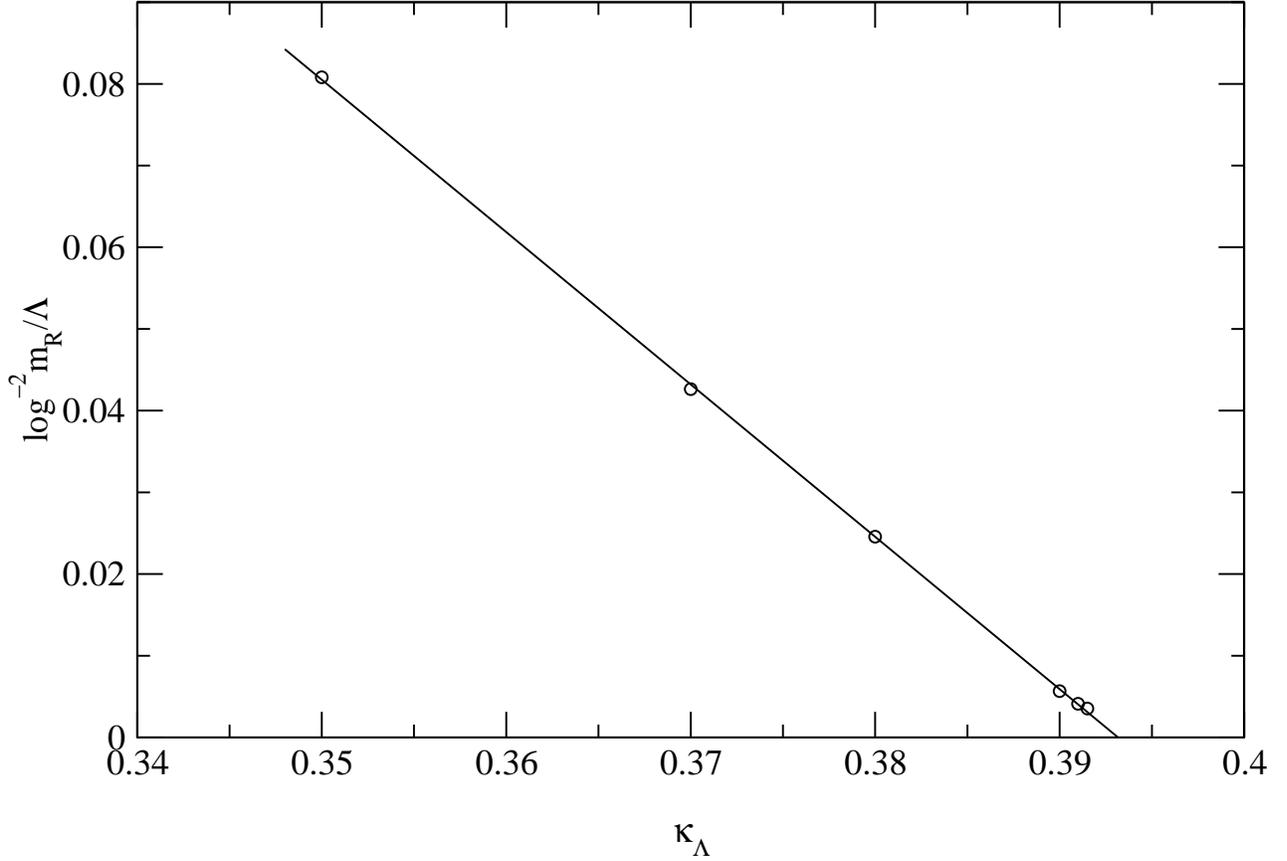}
\caption{Essential scaling for $d=2,\ N=2$. The renormalized mass $m_R$ is plotted as a function of $\kappa_\Lambda=\kappa_{\Lambda*}-H(T-T_c)$.}
\label{escfig}
\end{figure}
\begin{figure}[h]
\centering\epsfig{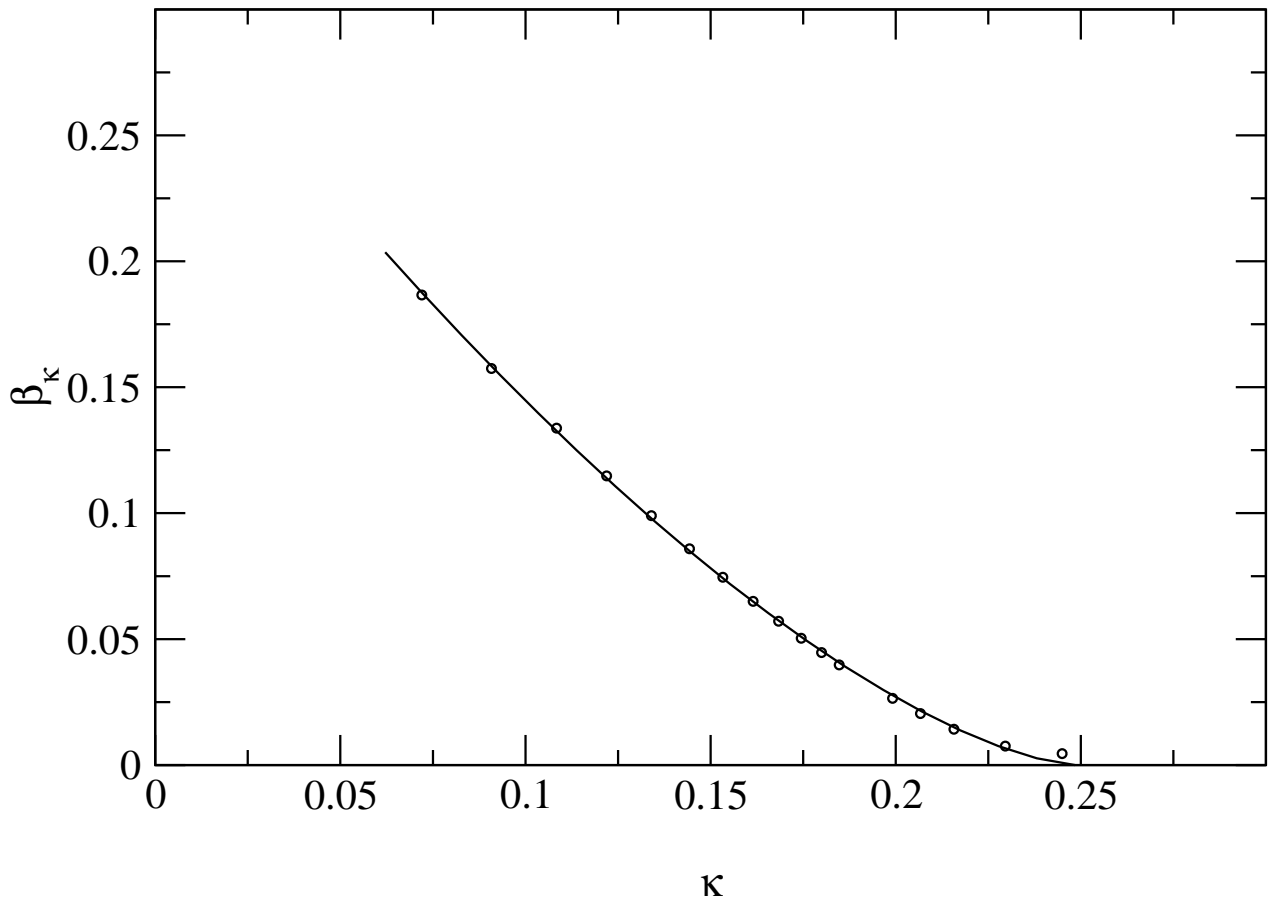}
\caption{Details of the beta function for $d=2$, $N=2$. The curve is a fit to eq. (\ref{betakappa param}).}
\label{esc2}
\end{figure}

The abelian case, $N=2$, is known to exhibit a special kind of phase transition which is usually described in terms of vortices \cite{KT73-1}. 
The characteristics of this transition are a massive high temperature phase and a low temperature phase with divergent correlation length but zero magnetization. The anomalous dimension $\eta$ depends on $T$ below $T_c$ and is zero above. It takes the exact value $\eta_*=0.25$ at the transition. The most distinguishing feature is essential scaling for the temperature dependence of $m_R$ just above $T_c$,
\begin{equation}
m_R\sim e^{-\frac{b}{(T-T_c)^\zeta}},\quad \zeta=\frac{1}{2}\label{esc}.
\end{equation}

\begin{figure}[h]
\centering\epsfig{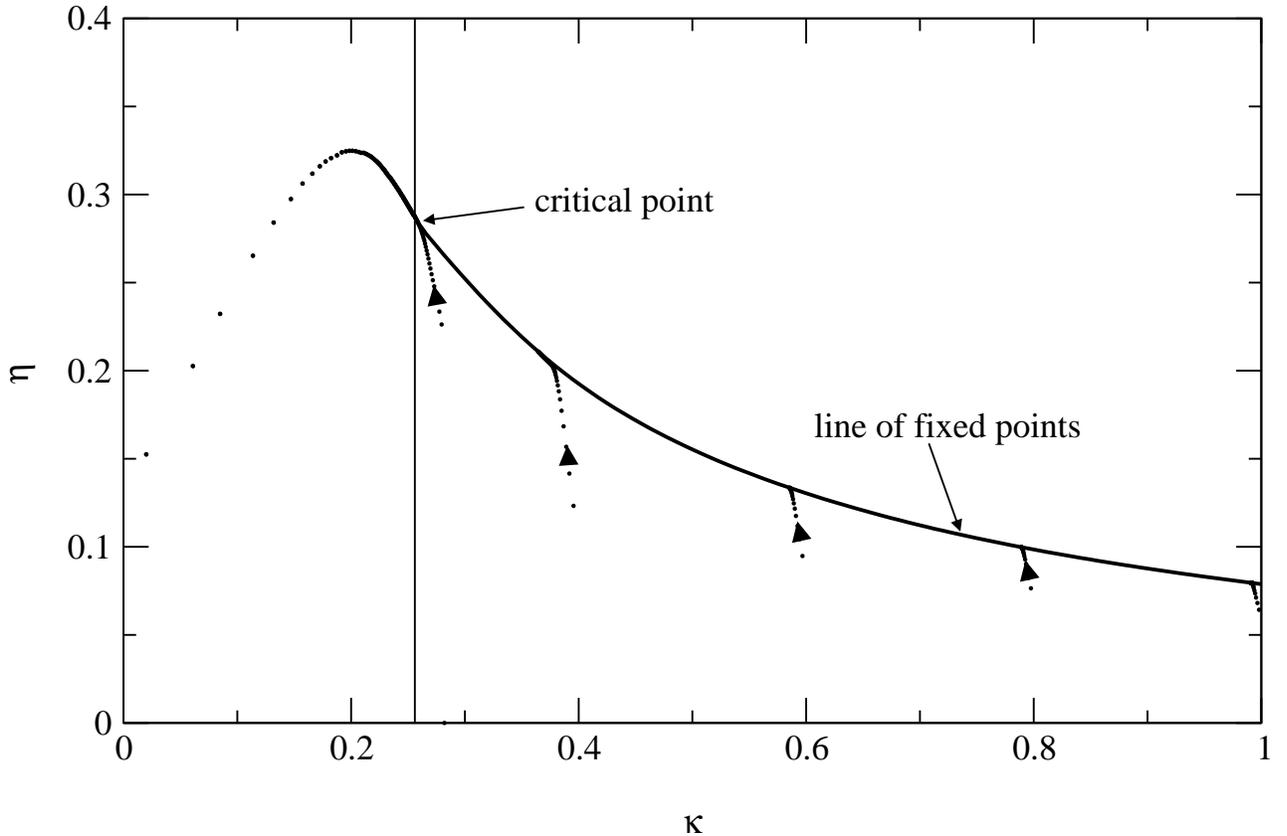}
\caption{Temperature dependence of the anomalous dimension $\eta$ for the low temperature phase, $d=2$, $N=2$. The line of fixed points is characterized by $\kappa$ and ends in the critical point for the Kosterlitz-Thouless phase transition. We also show the flow towards the line of fixed points and the flow in the high temperature phase away from the critical point (left). The spacing between the points indicates the speed of the flow.}
\label{etakappa}
\end{figure}

We have already mentioned the existence of a line of fixed points for large values of $\kappa$, which is relevant for the low temperature phase. The contribution of a massless (Goldstone) boson in the renormalization group equation ($w(k)=0$) is 
responsible for the finite value of $\eta$. This in turn drives the expectation value of the unrenormalized field to zero (even for a nonvanishing renormalized expectation value $\kappa$), 
\begin{equation}
\rho_0=\kappa/Z_k\sim\kappa(k/\Lambda)^\eta.
\end{equation}  
We can observe this line of fixed points to a good approximation (cf.\ fig.\ 2),
although the vanishing of $\beta_\kappa$ is not exact (we find $\beta_\kappa\approx -4\cdot10^{-5}\kappa^{-1}+{\cal O}(\kappa^{-2})$). 
Along the line of fixed points the anomalous dimension $\eta$
differs from zero. This reconciles the massless Goldstone boson
with Coleman's no to theorem \cite{Col73-1}
for a free propagator.
The line of fixed points ends at a phase transition which corresponds to a microscopic parameter $\kappa_{\Lambda*}$ as an initial value of the flow. 
In order to verify essential scaling we have to examine the flow for values of $\kappa_\Lambda$ just below that point, $\kappa_\Lambda=\kappa_{\Lambda*}+\delta\kappa_\Lambda$, $\delta\kappa_\Lambda\sim-(T-T_c)$. Then $\kappa_k$ crosses zero at the scale $k_s$ and we find the mass by continuing the flow in the symmetric regime (minimum at $\kappa=0$). In fig.\ 3 
we plot $(\ln{(m_R/\Lambda}))^{-2}$ against $\kappa_\Lambda$ and find excellent agreement with the straight line (\ref{esc}).

How does $\beta_\kappa$ have to look like in order to yield essential scaling? Since there is only one independent scale near the transition, one expects $m_R(T)=C_sk_s(T)$, where $k_s$ denotes the scale at which $\kappa$ vanishes, i.e. $\kappa(k_s,T)=0$. 
For $\kappa$ close to and below $\kappa_*$ we parameterize $\beta_\kappa$ (this approximation is not valid for very small $\kappa$)
\begin{equation}
\beta_\kappa=\frac{1}{\nu}\cdot(\kappa_*-\kappa)^{\zeta+1}.\label{betakappa param}
\end{equation}
For conventional scaling one expects $\zeta=0$ and the correlation length exponent is given by $\nu$. Integrating equation (\ref{betakappa param}) yields for $\zeta\neq0$, $\delta\kappa=\kappa-\kappa_*$:
\begin{equation}
\ln (k/\Lambda)=\frac{\nu}{\zeta}\left(\frac{1}{(-\delta\kappa)^\zeta}-\frac{1}{(-\delta\kappa_{\Lambda})^\zeta}\right)\label{integriert}.
\end{equation}
For $k=k_s$ the first term $\sim (-\delta\kappa)^{-\zeta}$ is small and independent of $T$ (since $-\delta\kappa(k_s)=\kappa_*$) and equation (\ref{integriert}) yields the essential scaling relation (\ref{esc}) for $\zeta=1/2$.
Usually, the microscopic theory is such that one does not start immediately in the vicinity of the critical point and the approximation (\ref{betakappa param}) is not valid for $k \approx \Lambda$. 
However, if one is near the critical temperature the trajectories will stay close to the critical one, $\kappa_c(t)$, with $\kappa_c(0)=\kappa_{\Lambda*}$. 
This critical trajectory converges rapidly to its asymptotic value $\kappa_*$ and $\beta_\kappa$ gets close to eq. (\ref{betakappa param}) at some scale $\Lambda'<\Lambda$. As a result, one may use equation (\ref{integriert}) only in its range of validity ($k<\Lambda'$) and observe that $\delta\kappa_{\Lambda'}$ is also proportional to $T_c-T$.
The numerical verification of (\ref{betakappa param}) is quite satisfactory: Fitting our data  yields $\kappa_*=0.248$, $\zeta=0.502$ and $\nu^{-1}=2.54$. The uncertainty for $\zeta$ is approximately $\pm 0.05$.
The numerical values of $\beta_\kappa$ and the approximation (\ref{betakappa param}) are shown in fig. 4.

\bigskip
\begin{center}
\begin{tabular}{|c|ll|ll|}
\hline
$N$
&\multicolumn{2}{c|}{$\nu$}
&\multicolumn{2}{c|}{$\eta$}
\\
\hline
0	&0.70	&0.75	&0.222	&0.2083...	\\
1	&0.92	&1	&0.295	&0.25\\
2	&--	&--	&0.287	&0.25\\
\hline
\end{tabular}
\end{center}

\medskip
\noindent Table 3\ 
{\em Critical exponents $\nu$ and $\eta$ for $d=2$. We compare each value with the exact result.}

\bigskip

One can use the information from figure \ref{escfig} or \ref{esc2} in order to determine $\kappa_*$ and therefore $\eta_*=\eta(\kappa_*)$, the anomalous dimension at the transition. We plot $\eta$ against $\kappa$ in fig.\ 5. 
 One reads off $\eta_*=0.287\pm0.007$ where the error reflects the two methods used to compute $\kappa_*$ and does not include the truncation error.
For $\kappa_\Lambda>\kappa_{\Lambda*}$ or $T<T_c$ the running of $\kappa$ essentially stops after a short ``initial running'' towards the line of fixed points. One can infer from figure \ref{etakappa} the temperature dependence of the critical exponent $\eta$ for the low temperature phase.
In summary all the relevant characteristic features of the Kosterlitz-Thouless transition are visible within our approach. 

We also have computed the critical exponents for the ``standard''
second-order phase transitions for $N=1$ (Ising model) and $N=0$.
They are displayed in table 3. The exponents are less accurate than the ones
for $d=3$ displayed  in table 1 ({\it g}). This confirms that the convergence
of the derivative expansion is controlled by the size of $\eta$.
Nevertheless, the agreement with the exact results is better than
10 \%.

We conclude that the first order in the derivative expansion of the exact flow equation for the effective average action gives a quantitatively accurate picture of all phase transitions of scalar models in the $O(N)$ universality class for arbitrary dimension $2\leq d\leq4$. As the next step,
a reliable error estimate would be very welcome.

\bigskip
\section*{Acknowledgement}
This review is based to a large extent on \cite{BTW96-1} and \cite{vGW}.
The author thanks J. Berges, G. von Gersdorff and N. Tetradis
for a fruitful collaboration.

\newpage

\end{document}